\documentclass
[
    a4paper,
    DIV=11,
    abstracton
]
{scrartcl}
\usepackage{amssymb}
\usepackage{amsmath}
\usepackage{fullpage,authblk,amssymb,amsthm,amsfonts,mathtools}
\usepackage{enumerate}
\usepackage{caption}

\usepackage{graphicx}
\usepackage[numbers]{natbib} 
\usepackage[normalem]{ulem}
\usepackage[usenames]{color}
\usepackage{xcolor}
\usepackage[colorlinks,
			pdffitwindow=false,
      plainpages=false,
      pdfpagelabels=true,
     	pdfpagemode=UseOutlines,
      pdfpagelayout=SinglePage,
			bookmarks=false,
			colorlinks=true,
      hyperfootnotes=false,
			linkcolor=blue,
			urlcolor = red!50!black,
			citecolor=green!50!black]{hyperref}

\graphicspath{{./Figures/}}

\newcommand\Pran{\mbox{\textit{Pr}}} 
\newcommand\Ra{\mbox{\textit{Ra}}} 
\newcommand\Ro{\mbox{\textit{Ro}}} 

\newcommand{\set}[1]{\mathbb{#1}}

\newcommand{\R}{\mathbb{R}}
\newcommand{\ep}{\varepsilon}

\newcommand{\N}{\mathbb{N}}

\newcommand{\im}[1]{\mbox{Im}\left({#1}\right)}
\newcommand{\imu}{\mathrm{i}}

\title{Diffusion maps embedding and transition matrix analysis of the large-scale flow structure in turbulent Rayleigh--B\'enard convection}
\author[1]{P\'eter Koltai}
\author[2]{Stephan Weiss}
\affil[1]{\small Department of Mathematics and Computer Science, Freie Universit\"at Berlin, Arnimallee 6, 14195 Berlin, Germany}
\affil[2]{\small Laboratory for Fluids, Pattern Formation and Biocomplexity (LFPB),
Max Planck Institute for Dynamics and Self-Organization,
Am Fassberg 17,
37077 Goettingen, Germany}
\date{}                                           

\begin{document}
\maketitle

\begin{abstract}
By utilizing diffusion maps embedding and transition matrix analysis we investigate sparse
temperature measurement time-series data from Rayleigh--B\'enard convection experiments in a
cylindrical container of aspect ratio $\Gamma=D/L=0.5$ between its diameter ($D$) and height ($L$). We consider the
two cases of a cylinder at rest and rotating around its cylinder axis.  We find that the relative
amplitude of the large-scale circulation (LSC) and its orientation inside the container at different
points in time are associated to prominent geometric features in the embedding space spanned by the
two dominant diffusion-maps eigenvectors. From this two-dimensional embedding we can measure
azimuthal drift and diffusion rates, as well as coherence times of the LSC. In
addition, we can distinguish from the data clearly the single roll state (SRS), when a single roll
extends through the whole cell, from the double roll state (DRS), when two counter-rotating rolls
are on top of each other. Based on this embedding we also build a transition matrix (a discrete
transfer operator), whose eigenvectors and eigenvalues reveal typical time scales for the stability
of the SRS and DRS as well as for the azimuthal drift velocity of the flow structures inside the
cylinder. Thus, the combination of nonlinear dimension reduction and dynamical systems tools
enables to gain insight into turbulent flows without relying on model assumptions.

\end{abstract}
\vspace{2pc}
%

\section{Introduction}

Due to technological advancements in past years the rate at which data can be acquired, stored and
processed has increased tremendously.
In addition, powerful computers allow to simulate natural processes in greater detail, with larger
temporal and spatial resolution than ever before. While acquiring data from measurements and
simulation is essential for scientific progress, the goal is always to develop simple
\emph{effective} models that map high-dimensional natural processes onto lower dimensional
models that, optimally, allow for predicting the future with accuracy.

Ideally, such a reduction results in functional relationships between just a few control and response parameters. Deriving equations that describe the system, however, can only be done if one
has a clear understanding of the underlying fundamental mechanisms. While such an understanding
often does not exist, even if the fundamental mechanisms are known, they are often not sufficient to
make accurate predictions in complex nonlinear systems. For example, while it is well known that the flow of
simple fluids is governed by the Navier--Stokes equations \cite{Davidson2009}, their possible
solution is computationally demanding and very sensitive to boundary and initial conditions, such that simple accurate
predictions are impossible.

Over the years many tools have been developed for the reduction of data without taking additional
knowledge about the source and type of these data into account.  Early pioneering approaches are
Principal Component Analysis (PCA) \cite{pearson1901liii,Jolliffe86,Shlens14}, where
multidimensional data are mapped on a lower dimensional space via a linear mapping by keeping as
much variation as possible, or Multidimensional Scaling (MDS)
\cite{kruskal1964multidimensional,TSL00}, where a linear embedding is  used such that pair-wise
distances between points are maintained as well as possible.  A more novel method, the {\em diffusion
maps approach}, first suggested by Lafon and Coifman \cite{Lafon04,CL06}, uses local information from
the data \cite{RoSa00,BeNi03,DoGr03,ZhZh04}, in contrast with the above methods that use global
distances.  It is a non-linear technique that parametrises data based on their underlying
connectivity, {\em i.e.,} their proximity in the space spanned by the observable variables. This
approach has been successfully used for instance for image analysis~\cite{BWWA13} and image
processing~\cite{FFL10}.

The combination of dimension reduction tools with dynamical time-series analysis can lead to a better
understanding of complex and high-dimensional systems: (i) PCA is applied in Proper Orthogonal
Decomposition (POD)~\cite{smith2005low,rowley2005model}; (ii) time-lagged correlation analysis
results in Dynamic Mode Decomposition~\cite{SS08,klus2018data}, which is a particular instance of
Koopman mode analysis~\cite{RMBSH09,williams2015data,KlKoSch16,arbabi2017study}, (iii) cluster
analysis~\cite{kaiser2014cluster} aggregates system states with respect to similarity, and (iv)
optimal transport is used to find a non-linear transformation that decouples the dynamics of
coordinates~\cite{ArSa19}. Data-based approximation of the Koopman operator by diffusion maps is
obtained in~\cite{BeGiHa15,giannakis2018koopman}, where the latter reference analyses a convection
problem similar to ours below, in a different container geometry. Classical embedding results from
differential geometry and dynamical systems with subdivision techniques are used
in~\cite{DeHeZi16,ZiDeGe18,GeKoDe19} to approximate finite-dimensional attractors and invariant
manifolds of infinite-dimensional systems.

In this work we propose to use the \emph{diffusion maps} embedding and \emph{transition matrix}
analysis~\cite{Hsu87,DeJu99} to analyse the structure and dynamics of a turbulent system. The main
advantage of our approach is that it represents the geometry of data in a low-dimensional space
where analysis techniques such as the transition matrix method can be applied. Thus, it
simultaneously delivers a geometric-topological and a dynamical understanding of the system.  The
particular system considered here is turbulent Rayleigh--B\'enard convection
(RBC). In RBC a horizontal fluid layer is confined by a warm plate from below and a cold plate from
the top. It is an archetypical system to study pattern formation and turbulence and has been investigated for more than a
century~\cite{Be00a,Rayleigh1916}. 

Under the Oberbeck--Boussinesq approximation \cite{Ob79,Bo03,SV60} the system is fully determined by
only two dimensionless parameters. These are the Rayleigh and Prandtl numbers,
\[
\Ra = \frac{g\alpha\Delta TL^3}{\kappa\nu} \quad\mathrm{and}\quad  \Pran=\nu/\kappa \mbox{},
\]
respectively. Here, $L$, $\Delta T$, and $g$ denote the height of the fluid layer, the temperature difference between its bottom and its top, and the gravitational acceleration. Furthermore, $\alpha$, $\nu$, and $\kappa$ are the isobaric thermal expansion coefficient, the kinematic viscosity and the thermal diffusivity.
The Rayleigh number measures the thermal driving, while the Prandtl number is the ratio of the two
damping mechanisms, {\em i.e.,} viscous and thermal diffusion. For small \Ra\ the flow is laminar,
and forms steady spatially periodic convection rolls with a wavelength of roughly twice the height
of the fluid layer. The convection flow becomes unsteady, chaotic and finally turbulent as \Ra\ increases.
In sufficiently extended systems, the signature of the initial periodic rolls can still be observed in the averaged
flow and temperature field as {\em turbulent superstructures}~\cite{PaSchSch18}.

For technical reasons, most experiments and numerical simulations have been conducted in smaller
cylindrical containers, with an aspect ratio $\Gamma=D/L$ between its diameter $D$ and its
height $L$ close to unity. In this case, the superstructure consists of a single 
large-scale circulation (LSC) roll, that extends over the entire cell, so that warm fluid
rises close to the sidewall on one side, while cold fluid sinks at the opposite side. 
This single roll structure is rather stable compared to the eddy turnover time, but exhibits interesting dynamics on
larger time scales, such as diffusive azimuthal drift, cessations, or torsional oscillations \cite{BA07a,FBA08,BA09}. 
The shape and dynamics of the LSC heavily depends on $\Gamma$. With increasing $\Gamma$ the single
roll state (SRS) is replaced by counter-rotating rolls arranged side-by-side \cite{PWBFH11}.

Well studied is the case of $\Gamma=1/2$.  Here the system is predominantly in the SRS, but the
single roll is significantly less stable than for $\Gamma=1$ and often undergoes a transition to a
state in which two counter-rotating rolls are on top of each other (double roll state -
DRS)~\cite{XX08,WA11c}. In the DRS, the vertical heat transport is reduced by about 
1\,-2\,\% compared to the SRS. Note that in the following, we refer to the large-scale flow as LSC
regardless of whether the system is in SRS or DRS.  Studying these large-scale coherent structures is
interesting as they pose forms of self-organisation in highly nonlinear systems very far from
equilibrium that are currently not understood at all. 

A particularly interesting variation of the classical RBC system is rotating RBC where the convection 
cylinder rotates around its vertical axis with a constant angular speed. Studying rotating
convection is crucial for a better understanding of the large convection system in geo- and astrophysics
that occur in rotating frames and are thus strongly influenced by Coriolis forces (see {\em e.g.,}~\cite{LE09,ZA10,KSNHA09,WWA16}). 

In this paper, we investigate the dynamics of the temperature field in turbulent RBC for the
rotating and non-rotating case using diffusion maps embedding. In particular, we re-analyse
temperature measurements at various locations in the sidewall of a convection cylinder of aspect
ratio $\Gamma=0.5$, filled with water at an average temperature of 40\,$^{\circ}$C. 
The analysis here was done with measurements at Rayleigh numbers of $\Ra=7\times 10^{10}$ and 9$\times 10^{10}$. 
Temperature measurements in the sidewall help to reveal large-scale convection structures, as in general, hot
fluid rises from the warm bottom plate along one side, while cold fluid sinks down from the top
plate along the opposite side (see fig.~\ref{fig:expSetup}a). Despite this large-scale circulation,
the turbulent fluid motion results in vigorous fluctuations of the temperature field in space and
time that are detected by the sidewall thermometers. The relevant data have been published before in
\cite{WA11a} for the non-rotating case and in \cite{WA11c} for the rotating case. 

In the next section, we will briefly describe the experimental setup used to acquire the data as
well as the structure of the data.  In section~\ref{sec:diffMap} we explain in detail the diffusion
map embedding and we will show the resulting embedding for a standard (horizontal and non-rotating)
RBC case. In section~\ref{sec:dynamics} we will analyse the embedded data regarding dynamical
features of the systems, and find that the long-term dynamical behavior can be connected to the
evolution of the LSC. 
We close the paper with a discussion section.

\section{Experimental setup and data collection}\label{sec:setup}
\begin{figure}[htpb]
\centering
\includegraphics[width=\textwidth]{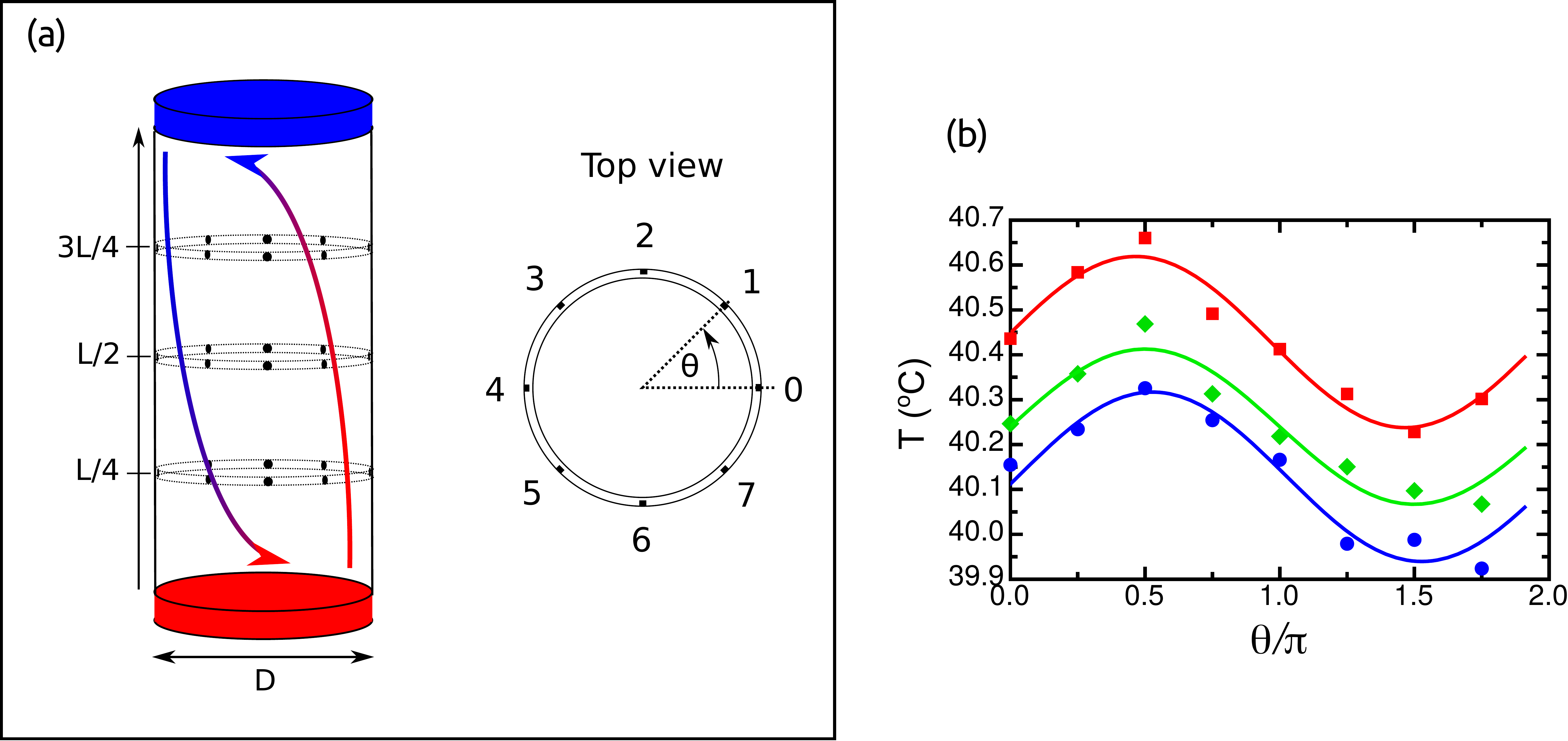}
\caption{(a) Sketch of the experimental setup with the location of the thermistors marked with
black dots (left) and cross-section of the cylinder with the azimuthal location of the thermistors
(right). Arrows mark the large-scale motion of the hot (red) and cold (blue) plumes that rise and
sink close to the sidewall. (b) Thermistor measurements at one time instance at the heights L/4 (red), L/2
(green) and 3L/4 (blue) showing the temperature signature of the large-scale circulation in the
single roll state. The solid lines are sinusoidal fits to the temperature as a function of the azimuthal position.
}
\label{fig:expSetup}
\end{figure}

A sketch of the experimental setup is shown in fig.~\ref{fig:expSetup}a. The main part of the
experiment is the cylindrical convection cell of aspect ratio $\Gamma=0.5$
that was closed by two horizontal copper plates from the bottom and the top. The top plate was
cooled using temperature regulated water, while the bottom one was heated by an ohmic heater. The
temperature of the bottom and top plate were kept constant to within $\pm 0.02$\,K of the desired
temperature during an experiment.

To characterise the temperature field, 24 thermistors were embedded in blind holes in the sidewall,
roughly a millimeter away from the fluid. Eight thermistors were equally distributed along the
azimuthal direction at each of the heights $L/4$, $L/2$, and $3L/4$. The working fluid was water
at a temperature of 40$^o$C, resulting in \Pran=4.38 for all analysed measurements in this
paper. Under the applied conditions the flow was highly turbulent with vigorous small
scale fluctuations that self-organised into a large-scale circulation that was predominantly in the SRS. 
The SRS can be detected in sidewall temperature measurements as 
sinusoidal temperature variation along the azimuthal direction at a specific height
(fig.~\ref{fig:expSetup}b).

During an experiment, the top and bottom plate temperature $T_t$ and $T_b$ were held constant and
the temperature of 24 thermistors embedded in the sidewall were recorded with a rate of
roughly one measurement every 3.4\,s. The experiment was conducted for several hours. Data
from the first hour were discarded as the system has not yet reached statistical equilibrium. 
Please see~\cite{WA11c} for further experimental details.

In the next sections we will reanalyse time series of the sidewall temperature measurements using
diffusion maps embedding. We will compare the results with results published in \cite{WA11c}.
The analysed datasets together with the relevant experimental parameters are listed in
table~\ref{tab:dataset}.

\begin{table}\begin{center}
\begin{tabular}{|l|c|c|c|c|}
\hline
Data Set: & No. of Data points & Ra                  & $\Omega$ [rad/sec] \\ \hline\hline
D1   & 302360             & 9.0$\times 10^{10}$ & 0.0                \\ \hline           
D2   & 9280               & 7.2$\times 10^{10}$ & 0.088              \\ \hline           
\end{tabular}
\end{center}
\caption{Experimental conditions for the two large datasets analysed in this paper.
}
\label{tab:dataset}
\end{table}

\section{Data embedding}\label{sec:diffMap}

In the following we consider the measurements of the 24 side wall thermistors as observations of the full system's
state (the full velocity and temperature fields) by a 24-dimensional observable at a given time. 
Before we start analyzing these data, we first describe the diffusion maps approach for an arbitrary data set (a cloud of data points). 

\subsection{Diffusion maps}
Let the set of measured data points~$\set{Z} := \{z_1,\ldots, z_m\} \subset \R^r$ be given. In our
case, the state of the system at a given time $t_j$ is represented by a 24-dimensional vector
($z_j\in \R^{24}$), spanned by the $r=24$ thermistor readings.
Note that the components of $z_j$ are dimensionless, as each component was calculated by subtracting
the mean temperature $(T_t+T_b)/2$ and normalising by the temperature difference $T_b-T_t$.
\footnote{The normalised temperatures $z_j$ relate to the measured temperatures $T_j$ as $z_j =
\frac{T_j-(T_b+T_t)/2}{T_b-T_t}$. Here $T_t$ and $T_b$ are the top and bottom plate temperatures and
$T_j$ is a 24-dimensional vector containing all temperature measurements at time $t_j$.}

The entire set $\set{Z}$ thus represents a cloud of all points taken during an experimental run,
i.e., $m=302360$ points in the 24-dimensional phase space for data set D1 (see table~\ref{tab:dataset}).
With the embedding algorithm explained below, we represent each state $z_j$ in a new coordinate
system, spanned by convenient and representative embedding coordinates~$\xi_{\bullet,j} := \xi_{\bullet}(z_j) \in \R^n$, $n<r$.  
Here $n$ is determined from the output of the diffusion maps algorithm and should be such that the
representation of the data set with these $n$ coordinates delivers insight into its geometry; in
general we use $n=2,3$ for visualization reasons. The \emph{reduced} embedding
coordinates~$\xi_{\bullet}$ will be a selected subset of the \emph{full} embedding
coordinates~$\xi$, as described below. We denote by~$\xi_i(z_j)$ the $i$-th (full) embedding
coordinate of the $j$-th data point, and $\xi_i$, the $i$-th coordinate function, can thus be
represented by a vector in~$\R^m$ through setting~$\xi_{i,j} = \smash{\left(\xi_i\right)_j} =
\xi_i(z_j)$, where~$(\cdot)_j$ denotes the $j$-th entry of a vector.

Note that the method will use only the vectors of measurements, and in particular it will use no
information whatsoever on the thermistor positions at which these measurements were taken. The
diffusion maps approach is designed to reveal the geometric features with the largest (nonlinear) variation in the data set it is applied to.

The method we choose to work with, \emph{diffusion maps}~\cite{CoLa06}, assumes that the given data
points are sampled from a low-dimensional smooth manifold that is embedded in the high-dimensional ambient space~$\R^r$. It will try to construct (non-linear) coordinate functions from the data set into a low-dimensional space that is one-to-one; thus giving a \emph{low-dimensional parametrization} (embedding) of the data set. 
To find this embedding map~$\xi_{\bullet}$ the method constructs a virtual diffusion process on the
data points, with the jump probabilities of this diffusion being based on proximity between 
data points---and thereby it neglects any natural or artificial ordering of the data, in particular any
temporal order. Such information, however, can later be re-included as we will do
below in section~\ref{sec:transmat}.

The guiding intuition behind the construction is that \emph{small local Euclidean} distances in the
ambient space $\R^r$ are good approximations of \emph{local geodesic} distances on the unknown data
manifold. This is not true for large distances, as the manifold might be curved. A
diffusion-like process
that is running locally ``along the data manifold'' thus ``feels'' the intrinsic geometry of this
manifold, and its properties thus reflect this geometry.

The construction is as follows. First we choose a \emph{proximity} or \emph{scale parameter} $\ep>0$, and define the data \emph{similarity matrix}~$K\in \R^{m\times m}$ by
\begin{equation} \label{eq:dmap1}
K_{ij} =  \exp\left(-\frac{\|z_i-z_j\|^2}{\ep}\right),\quad k_{\ep}(z_i) = \sum_{j=1}^m K_{ij},
\end{equation}
where algorithmically often a cutoff radius is used to set $K_{ij}\ll 1$ to zero, thus obtaining a sparse matrix with essentially no loss of accuracy. One sees, that $K_{ij}$ is a measure for the closeness between the two data points $z_i$ and
$z_j$ in their $r$-dimensional state space (i.e., the 24-dimensional space in our case). 
The row sums $k_{\ep}(\cdot)$ are now used to pre-normalize\footnote{The pre-normalization is necessary for canceling a bias of the data distribution in the results, if this distribution is non-uniform. For details, we kindly refer the reader to~\cite{CoLa06}, where this is done by introducing a tuning parameter~$\alpha$. Our construction is obtained with~$\alpha=1$.} the similarity matrix:
\begin{equation} \label{eq:dmap2}
\hat{K}_{ij} =  \frac{K_{ij}}{k_{\ep}(z_i) k_{\ep}(z_j)}, \quad d(z_i) = \sum_{j=1}^m \hat{K}_{ij}.
\end{equation}
By row-normalizing $\hat{K}$ we finally obtain the \emph{diffusion map matrix}
\begin{equation} \label{eq:dmap3}
P_{ij} = \frac{\hat{K}_{ij}}{d(z_i)} \mbox{.}
\end{equation}
With this normalization, the matrix elements $P_{ij}$ can be interpreted as a probability that a random walker moves
from $z_i$ to $z_j$.
$P$ is a stochastic matrix giving rise to a---by construction reversible---Markov chain; a virtual \emph{diffusion}
on the data points. The central observation in \cite{CoLa06} (as pioneered in related works
\cite{RoSa00,BeNi03,DoGr03,ZhZh04}) is that the right eigenvectors of $P$, denoted by $\Xi_i \in \R^m$, at
dominant eigenvalues $\Lambda_i \in \R$, can be used as intrinsic coordinates on the manifold.\footnote{Why this is the case, is discussed in Appendix~\ref{app:diffusion_distance}.} 
More precisely, if $(\Lambda_i, \Xi_i)$ denote eigenpairs of $P$, we embed the data point~$z_j$ into $\R^m$ by
\begin{equation}
\xi(z_j) = \left( \Lambda_1 \left(\Xi_1\right)_j,\ldots, \Lambda_m \left(\Xi_m\right)_j\right)^T \in
\R^m\mbox{.}
\end{equation}
Thus, the entry of the $i$-th eigenvector (scaled by the associated eigenvalue) at a data point is
used as $i$-th coordinate value for the embedded data point. As $m$ is usually large (e.g.,
$m=302360$ for dataset D1), this is not yet a simplification. However, if the data manifold is
low-dimensional, then only the first few eigenvectors carry geometric information, and the
subsequent ones are redundant. Thus, we choose $n$ of the dominant eigenmodes (which we denote, for notational simplicity, by the first $n$ subdominant ones, keeping in mind that we could skip some, e.g., taking the 2-nd, 4-th, and 7-th mode) 
to define the \emph{diffusion map}
\begin{equation} \label{eq:diff_embedding}
\xi_{\bullet}(z_j) = \left( \Lambda_2 \left(\Xi_2\right)_j,\ldots, \Lambda_{n+1} \left(\Xi_{n+1}\right)_j\right)^T \in \R^{n},
\end{equation}
where the main advantage is in $n\ll m$ and $n\ll r$. Note that we discarded the first eigenvector,
since by the row-stochasticity of $P$ we have $\Xi_1 = (1,\ldots,1)^T$ because $P \Xi_1 = \Xi_1$, and thus it is an entirely
non-informative coordinate. 
We summarize the notation used in this construction in Table~\ref{tab:dmap_notation}.
\begin{table}[h]
\renewcommand{\arraystretch}{1.5}
\centering
\begin{tabular}{l | c | l}
Description           & Object                                      & Properties / Notation \\ \hline
Measured point in &$z_j\in \set{Z} \subset \R^r$               & \\[-9pt]
physical phase space & & \\
$i$-th embedding coordinate & $\xi_i: \set{Z}\to\R \Leftrightarrow \xi_i \in \R^m$                            & $\xi_i =\Lambda_i\Xi_i$, where $P\Xi_i = \Lambda_i\Xi_i$ \\
Full embedding        & $\xi:\set{Z} \to \R^m$           & $\smash{\left(\xi(z_j)\right)_i} = \xi_i(z_j) = \xi_{i,j} = \smash{\left(\xi_i\right)_j}$ \\
Reduced embedding     & $\xi_{\bullet}:\set{Z} \to \R^n$ & $\xi_{\bullet,j} = \xi_{\bullet}(z_j)$
\end{tabular}
\caption{Summary of the notation used in the construction of the embedding by diffusion maps.}
\label{tab:dmap_notation}
\end{table}

The computation of pairwise distances $\|z_i-z_j\|$ as necessary in here, can be done efficiently up
to hundreds of dimensions utilizing k-d tree data structure~\cite{Fri18}, and subsequently solving a $m\times m$
sparse eigenvalue problem through vector iteration~\cite{Ste02}, as only the dominant modes are required. 
This latter step is usually the
computational bottleneck, and makes the method for $m \gg 10^4$ infeasible, depending on the
sparsity of~$P$. Fortunately, one can subsample the data, compute 
($\Lambda_i,\Xi_i$) pairs on this data set of
tractable size, and ``interpolate'' the embedding of the remaining points without the need of any
further eigenvalue computations. 
More information on how to do this, on how to choose the proximity
parameter $\ep$, and on why the diffusion map gives good intrinsic coordinates on the manifold
is deferred to Appendix~\ref{app:dmaps_appendix}.

\subsection{Applying the algorithm to the measurements}
\label{sec:application_to_data}

\paragraph{The data set.}

The measured data points are local observations~$z_j = h(x(t_j))$ at specific times $t_j$ by some observation function~$h:\set{X} \to \R^r$ of the original dynamics
\begin{equation} \label{eq:orig_dyn}
x(t_{j+1}) = F^{\tau}(x(t_j)),
\end{equation}
where $x(t)\in\set{X}$ is the full state of the system (the velocity and temperature field), and $F^{\tau}$ denotes the 
time dynamics of the system, {\em i.e.,} governed by the momentum and energy equations in our case. Although strictly not necessary, here we assume that the observation times $t_j$ are equispaced, i.e.,  $t_{j+1}-t_j = \tau$ for all~$j$.

In the experimental setup of section~\ref{sec:setup} the state $z_j\in\R^{r}$ is given by the
$r=24$ sidewall temperature measurements and thus marks a single point in a 24-dimensional parameter
space. The number of data points acquired during a single experimental run is usually at
least~$m\approx 10^4$.

We note that to uncover the attractor of a partially observed system, often delay-embedding in the
sense of Takens~\cite{takens1981detecting,robinson2005topological} is used. For a
noisy system, however, these approaches have their limitations, as one would necessarily reconstruct
the state space of the noise as well. Our analysis did not show a significant difference in the
geometry of the data set if analysed in delay-coordinates, thus in the following we will work in the
original 24-dimensional data space.

\paragraph{Geometric features of the data: small scales.}

We now calculate diffusion maps from sidewall data of dataset D1 that was acquired without
rotation of the convection cylinder. For this preliminary analysis we subsample the original data
set and take 3800 equally sampled points between times 3400\,s and 68930\,s, as measured from the beginning of the experiment.
We apply the automated procedure to find an appropriate scale parameter $\ep$, as described in
Appendix~\ref{app:opteps}. This automated procedure yields $\ep=1.25\cdot 10^{-4}$, and estimates a dimension of
the data manifold to be approximately~5. 
As we will see, the data does not have a simple manifold structure of locally homogeneous dimension
(at least not on this scale), and noise plays a role as well, thus this dimension estimate should
not be taken as a precise numerical value. Nevertheless, as the estimate is substantially smaller
than the ambient dimension 24, we can claim with high certainty that the data yields a
low-dimensional geometric structure.

\begin{figure}[htb]
\includegraphics[width=\textwidth]{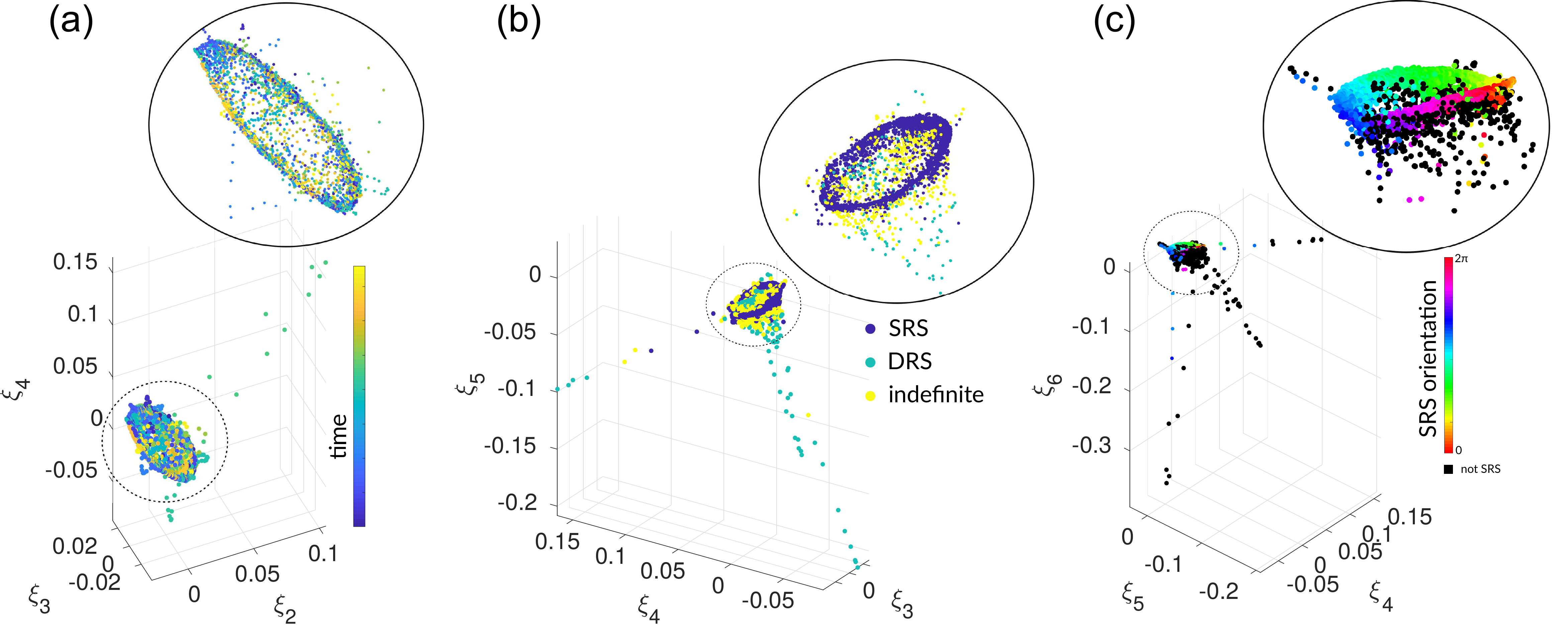}
\caption{Embeddings of the partial data set D1 with diffusion maps and proximity parameter $\ep = 1.25\cdot 10^{-4}$. 
Color code represents time (a), roll state of the LSC (b) and azimuthal orientation of the LSC (c);
please refer to main text for more details. Core regions in each embedding are magnified and shown
on top of each plot. 
}
\label{fig:1003261_smalleps}
\end{figure}

In fig.~\ref{fig:1003261_smalleps} we show three-dimensional embeddings $\xi_{\bullet}$ of the data
by $(\xi_2,\xi_3,\xi_4)$ (fig.~\ref{fig:1003261_smalleps}a), $(\xi_3,\xi_4,\xi_5)$
(fig.~\ref{fig:1003261_smalleps}b), and $(\xi_4,\xi_5,\xi_6)$ (fig.~\ref{fig:1003261_smalleps}c).
The coloring in this figure represents (a) time (time increases from blue to yellow),
(b) the roll state (dark blue: SRS, turquoise: DRS, yellow indefinite), and (c) the orientation of
the single roll (color hue: orientation, black: DRS and indefinite), as obtained in~\cite{WA11c}. We see
that all eigenvectors pick up a core disc-like structure and long excursions that depart from the
core and return to it shortly after. We will see below that the disc-like structure is a
representation of the large-scale flow structure in the system. The excursions on the other hand
might be interpreted as signatures of the intermittent nature of the turbulent flow. We already see
here that the disc-like core correlates with the amplitude (fig.~\ref{fig:1003261_smalleps} b) and the
orientation (fig.~\ref{fig:1003261_smalleps} c) of the LSC. This indicates that the low-dimensional
embedding readily represents important physical structures related the to large-scale circulation of
the convective flow---the \emph{turbulent superstructures} in convection---while the distinct
eigenvectors encode transient excursive behavior of different kinds. We note that this distinguishes
diffusion maps from linear methods, such as PCA, which are not able to effectively separate the bulk
of the data from the excursions (computations not shown here).

In this study we shall focus on the dynamically most prevalent structure, namely the core region
where most data points reside. Noting that the transients are rare events consisting of just a few
data points---which nevertheless have a large geometric deviation from the ``mean pattern''---, we
will consider an embedding that focuses on the large-scales and suppresses the transients. This is
achieved by increasing the proximity scale to $\ep = 5\cdot 10^{-4}$. An explanation of the
phenomenon is given via the diffusion distances in Appendix~\ref{app:diffusion_distance}. Further
increasing of $\ep$ would make the similarity of all point pairs asymptotically equal, thus we would
lose all the information on the geometric structure in the data. Taking $\ep \ll 1.25\cdot 10^{-4}$
would, contrariwise, disconnect all data points, resulting in a similar loss of information. We
refer to Appendix~\ref{app:opteps} for details. We also note that there is a reasonable robustness
of the method with respect to local variation of the proximity parameter.

\paragraph{Geometric features of the data: large scale.}

\begin{figure}[htpb]
\centering
\includegraphics[width=\textwidth]{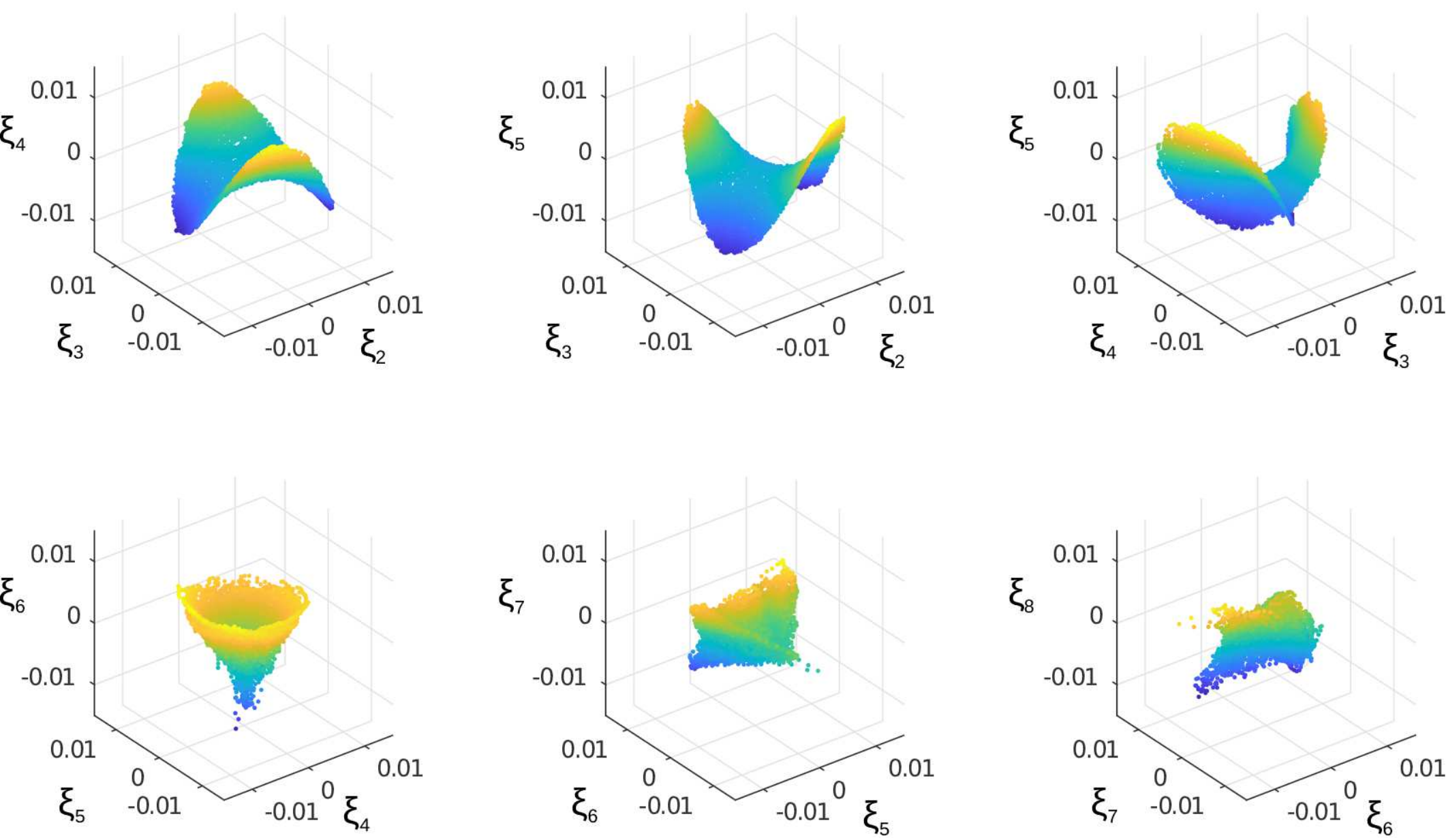}
\caption{Different diffusion map embeddings for the dataset D1. Color code marks the value of
the vertical coordinate for better visualisation. The kernel scale was~$\ep=5\cdot 10^{-4}$.}
\label{fig:1003261_pure}
\end{figure}

The results of the diffusion map embedding with $\ep=5\cdot 10^{-4}$, are shown in
fig.~\ref{fig:1003261_pure}, where we plot three-dimensional embeddings $(\xi_k, \xi_{k+1}, \xi_{k+2})$ for
$k=2,\ldots,6$ of the data set by the 8 eigenvectors with the largest eigenvalues. 
The corresponding eigenvalues are:  
\begin{table}[h]
\begin{center}
\begin{tabular}{|l|c|c|c|c|c|c|c|c|}
\hline
$\ep$ & $\Lambda_1$ & $\Lambda_2$ & $\Lambda_3$ & $\Lambda_4$ & $\Lambda_5$ & $\Lambda_6$ & $\Lambda_7$& $\Lambda_8$ \\ \hline\hline
$5\cdot 10^{-4}$ & 1.0000 & 0.7795 &0.7771 & 0.4938 & 0.4686 & 0.3356 & 0.2591 &  0.2445  \\ \hline
\end{tabular}
\end{center}
\end{table}

We see that the eigenvalues decrease sufficiently fast and thus we do not expect further
\emph{relevant} geometric information to be hidden in the lower spectrum. A comparison of the mutual
ratios of the log-eigenvalues $\log(\Lambda_i)$ with one another show similar ratios than the
eigenvalues of the Laplacian on a disc, indicating that the large-scale geometric structure of the
data set is a disc (see Appendix~\ref{app:Bessel}). Another evidence supporting this claim is that
the embedded points in fig.~\ref{fig:1003261_pure}(a-d) gather along a two-dimensional sub-manifold
in the 3d-space, meaning that two of the embedding coordinates parametrize the third one, hence this
does not yield additional geometric information. Moreover, these three-dimensional embeddings show
very strong similarities to analogous embeddings by the eigenfunctions of the Laplacian on a disc,
cf.~fig.~\ref{fig:LaplaceEmbed} in Appendix~\ref{app:Bessel}, indicating a disc-like data manifold.

We will thus focus in the following mainly on the embedding of the data in the
$(\xi_2,\xi_3)$-space. First, we shall compare the information obtained from this embedding with a
previous analysis of the data---which relies on the knowledge of the physical
setting of the underlying experiment, such as the shape of the container and the existence of
convection roll states.

\paragraph{Comparison with previous results.}

\begin{figure}[htpb]
\centering
\includegraphics[width=0.85\textwidth]{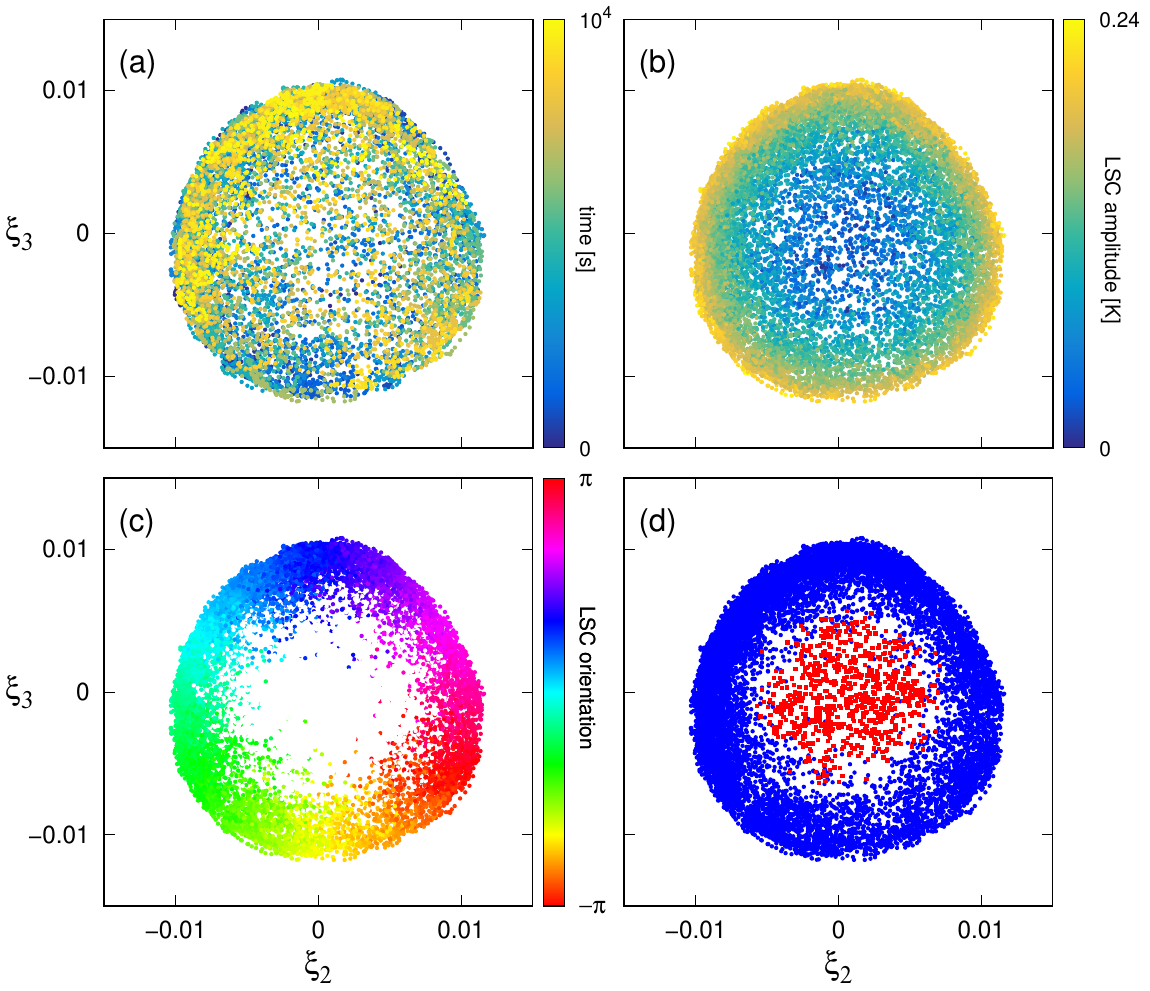}
\caption{Relation between physically determined properties of the large-scale circulation and
location of the states in the $(\xi_2,\xi_3)$ eigenvector space. Color code marks (a) time, (b)
amplitude of the LSC, and (c) orientation of the LSC as deduced from the azimuthal temperature
profile at midheight (color hue shows orientation).  Data points with very small amplitude have been
omitted in (c), since calculating the orientation for the LSC with very small amplitude is not only
ambiguous but also meaningless, as there might not be a well defined LSC.  Subfigure (d) shows data
points of the system when the large-scale circulation is in a single roll state (blue) or a double
roll state (red).}
\label{fig:embedding_12}
\end{figure}

Figure~\ref{fig:embedding_12} shows the $(\xi_2,\xi_3)$-embedding for~$\ep=5\cdot 10^{-4}$.
Additionally, some further properties obtained in~\cite{WA11c} of the physical system are
represented by the color of the data points.  In fig.~\ref{fig:embedding_12}a, colors represent time
at which the data were taken, ranging from blue at early times to yellow at later times. In this
representation no clear correlation between time and the location of the data in the eigenvector
space can be seen. That means, on sufficiently large time scales the system is in a statistically
steady state. Later, in section~\ref{sec:dynamics} we will further analyse the \emph{dynamical}
behaviour of the system.

The data points in fig.~\ref{fig:embedding_12}b are color-coded with the amplitude of the LSC. Here,
we see a clear correlation. Points with large radii $r_{\xi,j} := \smash{ \sqrt{\xi_{2,j}^2 + \xi_{3,j}^2}}$ also show large LSC 
amplitudes, while points inside the circle, with small radii also show small amplitudes.
Furthermore, as shown in fig.~\ref{fig:embedding_12}c, the angle $\theta :=\arctan{(\xi_3/\xi_2})$ also
correlates with the orientation of the LSC very well. We thus see that the two dominant eigenvectors describe the most prominent variation of the LSC, the large-scale motion of the system.

As we have pointed out already in the introduction, the LSC in cylinders of aspect ratio
$\Gamma=1/2$ can take the shape of a single roll (single roll state - SRS) or of two counter
rotating rolls (double roll state - DRS) one on top of the other.
In fig.~\ref{fig:embedding_12}d, we mark with blue dots whenever 
the system is in SRS, while red squares mark states, whenever the system is in DRS. We do not plot
any points for which the system is in a undefined state\footnote{Please see \cite{WA11a} for the detection criteria
of of both states.}. It can be seen that points corresponding to SRS and
DRS are clearly distinguishable. While SRS points are located on the outer areas of the disc, 
points corresponding to the DRS are located closer to the center of the disc. In this representation points belonging
to SRS and DRS are not well separated, and there is a large number of points that
belong to a transition state. This suggests that the DRS here is not a stable state but rather an
intermediated state after the SRS has been rendered unstable, e.g., due to the growth of a corner roll
(see \cite{SNSCZXSGXL10}). 

It is worth to stress again that diffusion maps finds the largest nonlinear geometric variation in
the data set, without any additional information about the system. Hence, while it identifies
structure in the data, it does not tell us \emph{what} these structures are \emph{in a physical
sense}, unless we provide additional knowledge, as for example about the positions of the
thermistors.  As an advantage, we can detect dominant dynamical behavior of the system without
any additional physical knowledge, as long as these are present in the identified coordinates~$\xi$.
This will be discussed in the next section.  In our case, a sufficiently spread-out arrangement of
the thermistors is vital for ``seeing'' the LSC in the data. If the thermistors would be located
rather close to each other, for example only at one half of the sidewall, the LSC could not have
been detected --- in this case the embedding would show other features related to the local
fluctuations of the temperature field at that particular location.

\section{Analysing dynamical features}\label{sec:dynamics}

It is in the nature of turbulent flows to show chaotic dynamics on different time and length scales. We 
therefore propose in the following new methods of statistically analysing the dynamical features of
the convective flow utilising the diffusion maps approach. The main focus will be on 
analysing our data in their $(\xi_2, \xi_3)$-embedding. Recall that we denote this reduced embedding of the $j$-th data point by~$\xi_{\bullet,j} := \xi_{\bullet}(z_j)$.

\subsection{The dynamics of the large-scale motion}\label{sec:orientdyn}

As we observe in fig.~\ref{fig:embedding_12}, the large-scale geometry of the dataset resembles a
disc with sparse occupation towards the center. In particular, data points that correspond to the
SRS form a ring. This suggests that the orientation on the ring (the ``angular coordinate'') is the
most important variable of the system, as it has the largest variation in the measurement space.
Thus, in the following we investigate the dynamics of this coordinate. Note that investigating the
stability and dynamics of the SRS has been a topic of interest for more than a decade
\cite{BA07a,BA08b,BA09,SBHT14}. 

To this end, for each embedded point~$\xi_{\bullet,j}\in\R^2$ we compute its azimuthal angle~$\theta_j =
\mathrm{arg}(\xi_{\bullet,j})$. Since we want to investigate the long time dynamics, we need to get rid of any
effects caused by its periodicity. Thus we unwind the data by assuming that within a single time steps the angle does not
change by more than $\pi$. Any larger changes are due to winding and thus we unwind $\theta_j$ by
adding or removing multiple of $2\pi$ until $-\pi<(\theta_j-\theta_{j-1})\leq \pi$.

Now, for a given offset $s\in\N$, we calculate $\Delta \theta_j(s) = \theta_{j+s}-\theta_j$ and from
this the corresponding mean displacement $\langle \Delta \theta_j(s)\rangle$ and its variance
$\langle \Delta\theta_j(s)^2 \rangle - \langle \Delta\theta_j(s)\rangle^2 $. Here the average
$\langle\cdot\rangle$ is done over all times, i.e., over all $j$.

\paragraph{Non-rotating convection cylinder.}

\begin{figure}[h]
\centering
 \includegraphics[width = 1\textwidth]{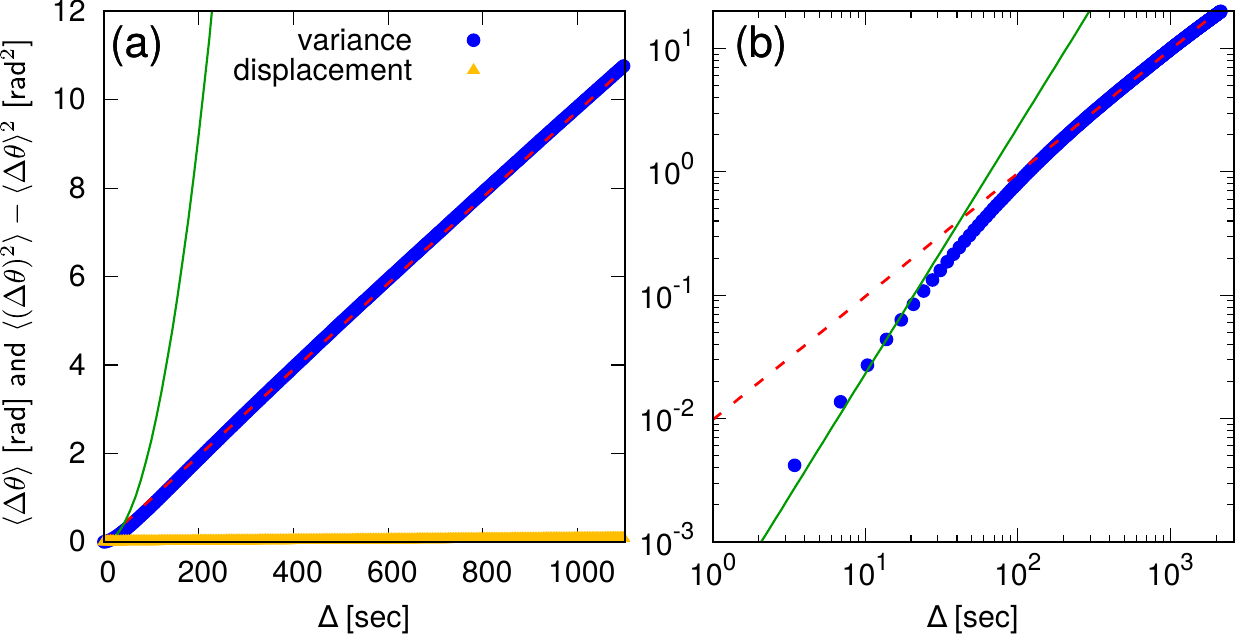}
\caption{Mean displacement (yellow triangles) and variance (blue bullets) plotted as a function of the lagtime $\Delta$ for the case without
rotation (data set: D1). Diffusion maps were calculated using $\ep = 5\cdot 10^{-4}$. (a): linear scaled
axis, (b): double-logarithmic axis. The green solid line marks a parabolic fit to variance for $\Delta<30$\,s
(coefficient $a=(2.3\pm 0.2)\times 10^{-4}$\,rad$^2$/s$^2$). The red dashed line marks a linear fit to the
variance resulting in a coefficient $D=(9.74\pm 0.01)\times10^{-3}$\,rad$^2$/s. A fit to the drift reveals an
average drift of $\omega_0=(7.2\pm 0.005)\times 10^{-5}$\,rad/s (fit not shown in the plot).}
\label{fig:1003261_orientdyn}
\end{figure}

For experiment D1 with $\ep = 5\cdot 10^{-4}$, we analyse the data for the $(\xi_2, \xi_3)$ embedding and
show the result in fig.~\ref{fig:1003261_orientdyn}. Fig.~\ref{fig:1003261_orientdyn}a shows the
mean displacement (drift) and variance as a function of the lagtime $\Delta := \tau s$ on a linear scale, whereas the same data are plotted on the double-log scale in~(b).

We see in fig.~\ref{fig:1003261_orientdyn}a that the average displacement (yellow triangles) is
negligible compared to the variance. This is expected, since there is no source for a constant mean
drift of $\theta$ with time. The mean drift is not exactly zero as we average over a finite number
of samples (i.e., short sections in time) and would further decrease for longer time traces.  We
will see below that the same system exhibits a nearly constant drift when the convection cylinder is
rotated around its vertical axis with a constant rotation rate.

The variance (blue bullets in fig.~\ref{fig:1003261_orientdyn}) increases monotonically with
increasing $\Delta$. From fig.~\ref{fig:1003261_orientdyn}b we see that the increase appears to be linear for 
$\Delta\gtrsim 100 $, suggesting a diffusive process on large time scales. A fit to the data reveals a 
diffusion coefficient (i.e., the slope of the curve) of $D=(9.74\pm 0.01)\times10^{-3}$\,rad$^2$/s. 
For small lag-times ($\Delta<50$\,s) the variance does not follow a linear trend. Its slope in the 
log-log plot (fig.~\ref{fig:1003261_orientdyn}b) is close to 2, which 
suggests a ballistic behaviour, i.e., $ \langle(\Delta\theta)^2\rangle
-\langle\Delta\theta\rangle^2\propto at^2$ for some $a>0$. Such dynamics, {\em i.e.,} a ballistic
motion for small time scales and a diffusive behaviour for large time scales is characteristic for
example for molecular motion and can be well described by a Langevin equation
\begin{eqnarray}
\label{eq:langevin}
\frac{d\theta(t)}{dt} &=& \omega(t) + \omega_0\\
\frac{d\omega(t)}{dt} &=& -\gamma \omega(t) + \eta(t)\mbox{.}
\end{eqnarray}
Here, $\omega_0$ is a constant angular drift that might occur, for example in a rotating frame and
$\gamma$ is a resistance caused by friction. The stochastic noise $\eta(t)$ describes contributions of turbulent fluctuations to 
the dynamics and is assumed to be $\delta$-correlated, with $\langle \eta(t_1)\eta(t_2)
\rangle = \sigma^2 \delta(t_1-t_2)$. These fluctuations lead to increments of the form $\eta(t)dt =\sigma dW(t)$, with $dW(t)$ being a standard Wiener process.

For eq.~(\ref{eq:langevin}) the mean displacement and its variance can be calculated~\cite{Gardiner04}, 
giving
\begin{equation}
\langle \Delta\theta(t)\rangle = \omega_0 t \mbox{,}
\end{equation}
and
\begin{equation}
\renewcommand{\arraystretch}{1.75}
 \langle \Delta\theta^2\rangle - \langle\Delta\theta\rangle^2 \propto \left\{
\begin{array}{ll}
at^2 := \frac{\sigma^2}{2\gamma}t^2, & t\to 0, \\
Dt := \frac{\sigma^2}{\gamma^2}t, & t\to\infty.
\end{array}
\right.
\label{eq:langevin_trends}
\end{equation}

Thus, the drag and forcing coefficients $\gamma$ and $\sigma^2$ can be computed from the diffusion
coefficient $D$ and the coefficient $a$ of the parabola, yielding 
$\gamma = 2a/D = (4.7\pm 0.4)\times 10^{-2}$\,s$^{-1}$ and $\sigma^2 = \gamma^2D = (2.2\pm 0.4)\times 10^{-5}$\,rad$^2$/s$^3$. 

A Langevin equation has already been suggested for the orientation of the LSC in
turbulent thermal convection in cylinders with aspect ratio $\Gamma=1$ by Brown and
Ahlers~\cite{BA08a}. 
In order to compare our results with their results, we have to compensate for the difference in
Rayleigh number. This can be done, as our cell height was similar to theirs and they provide
fitted power law relations between \Ra\ and $D_\theta$. While the fits in \cite{BA08a} and thus our estimate
includes significant uncertainty, we estimate a diffusion coefficient of $D_{\theta}=0.8\times 10^{-3}$\,rad$^2$/s$^3$, 
which is more than an order of magnitude smaller than for our case. This result 
reflects the larger fluctuations of the flow and smaller stability of the LSC in $\Gamma=1/2$
cylinders.

\paragraph{Rotating convection cylinder.}

\begin{figure}[h]
\centering
  \includegraphics[width = 1\textwidth]{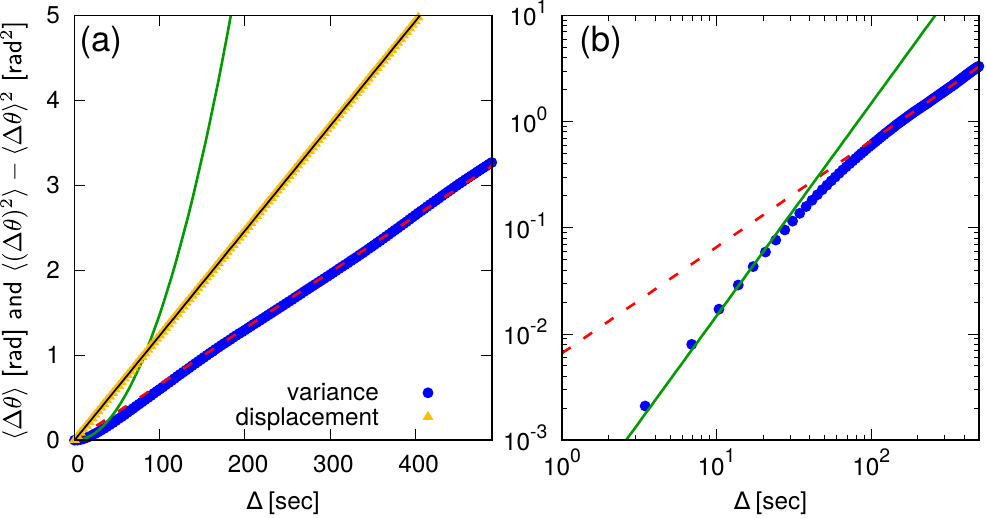}
\caption{Mean displacement (yellow triangles) and variance (blue bullets) as a function of the lagtime $\Delta$ for the
rotating case (dataset D2). (a) shows the data plotted against linear scaled axis. (b) shows
only the variance plotted against logarithmic axis. The lines are fits to the data. Black: linear
fit to the drift for $\Delta<500$\,s resulting in a slope $\omega_0 = (1.22\times 10^{-2})$\,rad/s. Green: Parabolic
fit to the variance for $\Delta<20$\,s resulting in a coefficient $a=(1.48\pm 0.02)\times 10^{-4}
$\,rad$^2$/s$^2$. Red dashed:
Linear fit to the variance for $100$\,s$<\Delta<500$\,s resulting in a diffusion coefficient
$D=(6.59\pm 0.01)\times 10^{-3}$\,rad$^2$/s.}
\label{fig:1102271_orientdyn}
\end{figure}

While for the non-rotating case, discussed above, the net drift was very small and just a
statistical feature that would shrink even more with longer time series, the drift becomes
significant when the cylinder is rotated around its cylinder axis.
Figure~\ref{fig:1102271_orientdyn} shows an analysis of the embedded data acquired in a cylinder
under rotation with a rotation rate of 0.088\,rad/s (data set~D2). As a result due to Coriolis forces the internal convection structure also rotates with respect to the side walls and thus with respect to the temperature probes deployed in them. 

We see that now $\Delta\theta$ increases linearly with $\Delta$. A linear fit to the date reveals a slope of
$\omega_0=(1222\pm 0.01 )\times 10^{-5})$\,rad/s. The variance of $\Delta \theta$ on the other hand
looks very similar to the non-rotating case in fig.~\ref{fig:1003261_orientdyn}. 
The variance increases initially ($\Delta\lesssim 30$\,s) quadratically with a coefficient 
$a=(1.48\pm 0.02)\times 10^{-4}$\,rad$^2$/s$^2$ and for larger $\Delta$ linearly with a diffusion
coefficient $D=(6.59\pm 0.01)\times 10^{-3}$\,rad$^2$/s, corresponding to $\gamma = (4.49\pm 0.06)\times 10^{-2}$\,s$^{-1}$
and $\sigma^2=(1.33\pm 0.04)\times 10^{-5}$\,rad$^2$/s$^3$. 

It is quite interesting that while the drag coefficient $\gamma$ is very similar to the non-rotating
case, $\sigma^2$ and thus the diffusion coefficient are significantly reduced, by almost 40\%.  This
shows how slow rotation suppresses turbulent fluctuations, resulting in a much more stable large
scale circulation. This stabilising effect has also been observed in other statistical quantities
such as the frequency of transitions between the double and the single roll state, the width of the
probability density function of the LSC amplitude, or the number of Fourier modes determined from
sidewall measurements~\cite{WA11d}.

We note that \Ra\ was roughly 15\% smaller for the rotating data set (D2) as for the non-rotating
one (D1). The difference in \Ra\ is too small to have any significant influence on either $\gamma$
or $D$, as we have observed from analysing other non-rotating data sets with $\Ra=7.2\times 10^{10}$
that were however much shorter and thus less suitable for a rigorous statistical analysis.

\paragraph{A note on the radius evolution.}

We have seen in section~\ref{sec:application_to_data} that the radius $r_\xi =\sqrt{\xi_2^2+\xi_3^2}$ corresponds 
to the amplitude of the LSC. We therefore also want to analyse the stochastic behaviour of $r_\xi$. 
Similarly to $\theta$, we now look at displacements $\Delta r_\xi(\Delta)$ for given lagtimes $\Delta$ and 
calculate their variance $v_{r}:=\langle(\Delta r_\xi)^2\rangle - \langle \Delta r_\xi\rangle^2$.

\begin{figure}[h]
\centering
  \includegraphics[width = 0.49\textwidth]{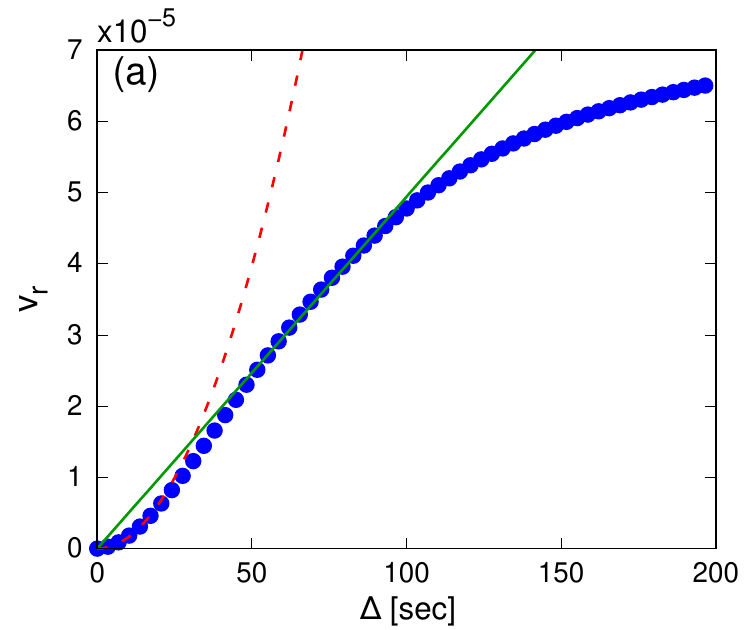}
  \includegraphics[width = 0.49\textwidth]{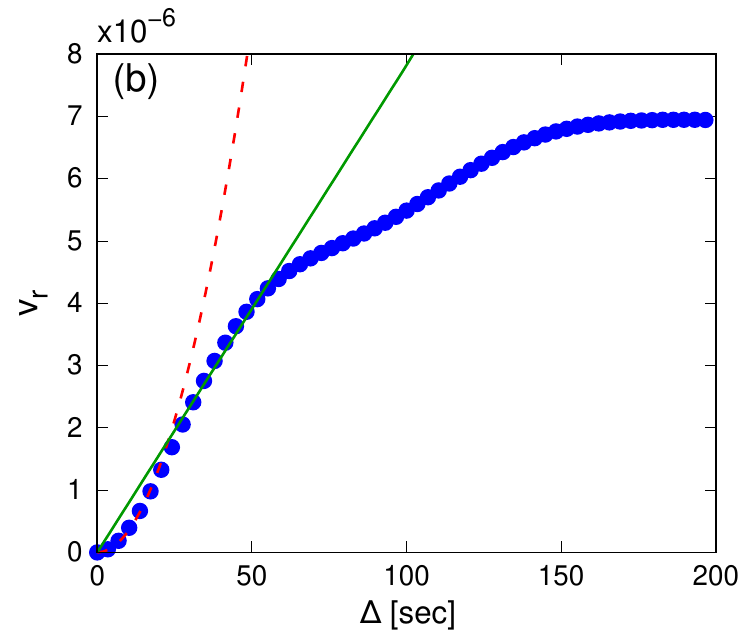}
\caption{Variance $v_r$ of the change of the radius $\Delta r_{\xi}$ as a function
of lagtime $\Delta$ (blue bullets).
Subplot (a) shows the non-rotating case (dataset D1). The red line marks a parabolic fit for
$\Delta<20$\,s with coefficient $a=(1.59\pm 0.03)\times 10^{-8}$ and the green line marks a linear
fit to the data with 50\,s$<\Delta<100$\,s, resulting in a slope of $D=(4.94\pm 0.02)\times 10^{-7}$.
Subplot (b) shows the rotating case (dataset D2). The red line marks a parabolic fit for
$\Delta<20$\,s with coefficient $a=(3.40\pm 0.07)\times 10^{-9}$ and the green line marks a linear
fit to the data with 40\,s$<\Delta<60 $\,s, resulting in a slope of $D=(7.8\pm 0.01)\times 10^{-8}$.
The $\Delta$-range for which we assumed a linear behaviour was chosen by eye.
}
\label{fig:rrms}
\end{figure}

We plot in fig.~\ref{fig:rrms} $v_r$ as a function of $\Delta$. For the non-rotating case
(fig.~\ref{fig:rrms}a) we see for $\Delta<20$ also a ballistic regime, where data follow a parabola
with coefficient $a=(1.59\pm 0.03)\times 10^{-8}$. For larger $\Delta$ there is a short range where
the data follow a linear function of ~$\Delta$. The corresponding slope is $D=(4.94\pm 0.02)\times 
10^{-7}$. For even larger $\Delta$, the slope of the data decreases again. This decrease is
expected, as $r_\xi$ can not take arbitrarily large values as also the amplitude of the LSC is 
confined. We note that Brown and Ahlers~\cite{BA08a} have modeled the dynamics of the amplitude of
the LSC as a Brownian motion inside a potential well. 
As fitting the complete model would be more involved, we defer this to future work. With the more general method presented in the next section, the statistical behaviour of $r_\xi$ is captured as well.
Note, that information about the exact amplitude of the LSC can not be calculated from $r_\xi$ 
but only qualitative features of it. 

Fig.~\ref{fig:rrms}b shows a very similar analysis for the 
rotating case (dataset D2). There, the coefficients $a$ and $D$ are significantly smaller, 
suggesting that also in this quantity the stabilising effect of rotation can be seen. 
We do not elaborate on this finding any further, as clearly a simple Langevin model
(eg.~(\ref{eq:langevin})) is only valid for
very small deviations, and $v_r$ is a nonlinear function of $\Delta$, just as the relationship between $r_\xi$
and the amplitude of the LSC is assumed to be nonlinear.
Furthermore, the relationship between the LSC amplitude and $r_\xi$ might be even different for the non-rotating and the rotating case.

\subsection{Transition matrix}
\label{sec:transmat}

In the previous subsection we have discussed mainly the angular dynamics of the single convection
roll (SRS) by considering the dynamics of the polar coordinate in the $(\xi_2,\xi_3)$ parameter
space. However, the dynamics could have other important features not readily revealed by the
embedding geometry. In examples below we will investigate the switching between SRS and non-SRS, and
also the dynamics observed for a small-$\ep$ embedding in three dimensions, where the transient excursions of the dynamics are still dominating the geometry.

The tool for this is going to be the transition matrix, which describes the redistribution of states
under the dynamics between subsets of the state space. As such, it is an approximation of the
so-called \emph{transfer operator} that describes the evolution of distributions due to the dynamics
of the system~\cite{LaMa94,DeJu99}.

More precisely, we consider the $m$ embedded data points in an $n$-dimensional
eigenvector space, {\em i.e.,}  $\set{Y}:=\{\xi_{\bullet,1},\ldots,\xi_{\bullet,m} \} \subset \R^n$.
Let~$\mathcal{P} = \{\set{B}_1,\ldots, \set{B}_p\}$ be a \emph{partition}\footnote{Usually the $\set{B}_j$ are non-overlapping axis-parallel $n$-dimensional ``boxes''; that is, $\set{B}_b = I_{b,1} \times \ldots \times I_{b,n}$, for some intervals $I_{b,i}$, $i=1,\ldots,n$.} of a domain~$\set{D}\subset\R^n$ (i.e., $\smash{\set{D} = \bigcup_{b=1}^p \set{B}_b}$), such that $\set{D}$ covers~$\set{Y}$ (i.e., $\set{Y} \subset \set{D}$), and every box is necessary for the covering, i.e., $\set{B}_b \cap \set{Y} \neq \emptyset$ for every~$b=1,\ldots,p$.

Recall that the data points were obtained from a long simulation, and were sampled at a nearly
constant rate with time period~$\tau = 3.4$\,s. Thus, the ordered time series $(\xi_{\bullet,j})_{j=1,\ldots,m}$, 
represents the dynamical time series $\smash{\big(x(t_j)\big)_{j=1,\ldots,m}}$ observed
through~$\xi_{\bullet}\circ h$. Consequently, one can view $\xi_{\bullet,j+s}$ as 
the image of~$\xi_{\bullet,j}$ under the dynamics after a lagtime $\Delta = s\tau$, where $s$ is the chosen \emph{offset}.
Let~$\set{J} = \{1,\ldots,m-s\}$. The discretization of the dynamics is then done by constructing the so-called transition matrix $T = T(s)\in \R^{p\times p}$ by
\begin{eqnarray} 
\label{eq:transmat}
T_{i\ell} &=& \frac{ \#\{j\in \set{J}:\, \xi_{\bullet,j}\in \set{B}_i \mbox{ and } \xi_{\bullet,j+s} \in \set{B}_\ell \} }
                   { \#\{j\in \set{J}:\, \xi_{\bullet,j} \in  \set{B}_i\} }\\ \nonumber
		  &\approx& \mathrm{Prob}\left[ \xi_{\bullet}(z(\Delta)) \in \set{B}_{\ell}\,\big\vert\,  \xi_{\bullet}(z(0))\in \set{B}_i \right],
\end{eqnarray}
counting the relative number of transitions from box $i$ to box~$\ell$, within the time from step
$j$ to $j+s$. Note that we can use an arbitrary offset~$s\in\N$ in~(\ref{eq:transmat}), and arbitrary partition $\mathcal{P}$ of the data set, but should consider the following aspects such that the transition matrix represents the dynamical properties of the system well~\cite{prinz2011markov,schutte2011markov}:
\begin{enumerate}[(a)]
\item
The offset should not be too small with respect to the size of the boxes $\set{B}_i$, otherwise most points do not leave the box they start in, and thus the dynamical features remain invisible.
\item
The offset should not be too large with respect to the number of data points in the boxes, otherwise the Monte Carlo estimate~(\ref{eq:transmat}) might be strongly erroneous.
\item
On a similar note, the boxes should not be too small  with respect to the data density, otherwise there are too few data points per box, and again one obtains too high sampling errors.
\end{enumerate}

By construction, $T$ is a row-stochastic matrix. Its elements approximate the probabilities that a
point from a certain box $\set{B}_i$ is mapped into another box $\set{B}_\ell$ by the dynamics. For determining the box partition and assembling the transition matrix we use the MATLAB-based software package 
\texttt{GAIO}~\cite{dellnitz2001algorithms}; cf.~\href{https://github.com/gaioguy/GAIO/}{https://github.com/gaioguy/GAIO/}.

Note that in principle this analysis can be applied directly to the 24-dimensional parameter space
in which our original data points $x_k$ are observed. However, partitioning a high-dimensional space
suffers from the curse of dimensionality, and becomes quickly computationally intractable. 
Also, with the diffusion maps embedding we can represent our data points based on a few chosen most
important independent features and neglect everything else. In the next section we will elaborate
how spectral analysis of $T$ can uncover the long-term dominant dynamical behavior of the underlying
process.

Assembling the transition matrix by~(\ref{eq:transmat}) scales linearly with the number of data
points, making it numerically tractable for a large amount of dynamical data. Because for diffusion
maps the computational bottleneck is to acquire the pairwise distances between (close-by) data
points and to solve the eigenvalue problem, it would be desirable to do this computation on data
sets with~$m \lesssim 10^4$. These seemingly opposing attributes can be brought together by the
\emph{out-of-sample extension} of diffusion maps~\cite{CSSS08}. On the one hand, for a
representative subset $\set{Y}^\prime \subset \set{Y}$ of data points the diffusion maps embedding
of the dominant geometric features of the data set is not improved by increasing the size
of~$\set{Y}^\prime$. On the other hand, having computed a diffusion map embedding for some
$\set{Y}^\prime$, there is a substantially cheaper way to approximate the embedding coordinates of
additional points (i.e.,~$\set{Y}\setminus \set{Y}^\prime$) than to compute the embedding for the
full data set. This method is described in Appendix~\ref{app:dmaps_interp} and it is utilized in the
following to compute the embedding on less than $5\times 10^3$ data points, and to ``interpolate''
the embedding of up to $3\times 10^5$ additional data points. These are then used to calculate the transition matrix~$T$. 

\subsection{Extracting dynamical features from the transition matrix}

\paragraph{Stationary distribution.}
The transition matrix provides access to the long-time behavior of the dynamics. For instance, a
stochastic dynamics governed by the matrix $T$ (i.e., a Markov chain jumping from box to box) will
be ergodic if $T$ is irreducible, {\em i.e.,} every box can be reached from any other box with a
positive probability, thus $T^k_{i\ell}>0$ for a sufficiently large~$k$. 
Then the relative time duration that the process spends in box $\set{B}_b$ is given by the $b$-th component of the \emph{stationary distribution}~$\mu = (\mu_1,\ldots,\mu_p)$ that is given by~$\mu^TT = \mu^T$ with~$\sum_{a=1}^p \mu_a=1$. It is straightforward to see that for our transition matrix holds
\[
\mu_b = \frac{\#\{j\in \set{J}:\, \xi_{\bullet,j} \in  \set{B}_b\}}{\#\set{J}}.
\]
$\mu_b$ represents the probability that a random data point lies within box $\set{B}_b$ and it is~$\mu_b=\lim_{k\rightarrow \infty} T^k_{ib}$. 

\paragraph{Almost invariant behavior.}

If the transition matrix was not irreducible, there would be disjoint sets $\set{I}_1 \cup \ldots \cup \set{I}_\ell  = \{1,\ldots,p\}$ such that $T_{ab} = 0$ whenever $a\in \set{I}_i$ and $b\in
\set{I}_j$, $i\neq j$. The index sets $\set{I}_i$, and equivalently the sets~$\smash{\bigcup_{b\in\set{I}_i}\set{B}_b}$, are then called invariant. 
This means that there are different sets of boxes, between which transitions are impossible. A point $\xi_{\bullet,j}$ can only transfer to other boxes that are part of its own box set, but not to boxes in other sets.   One can show that in this case $T$ has an $\ell$-fold eigenvalue~1. There would be right eigenvectors $\rho_k$ satisfying
\begin{equation} \label{eq:invariant_evec}
\big(\rho_k\big)_i = \left\{ \begin{array}{ll}
1, & i\in \set{I}_k, \\
0, & i\notin \set{I}_k.
\end{array}
\right.
\end{equation}
Whether the transition matrix is irreducible or not highly depends on the time duration considered here, {\em i.e.,} the offset $s$. For example, for $s=0$ every box is decoupled from the other boxes, while for $s\rightarrow \infty$ we gain the stationary solution $T_{ab}=\mu_b$ (naturally, one would also need an arbitrarily long data set to be able to set up $T$). 

Now, if for some~$s$ the transition matrix is not irreducible but close to an irreducible matrix (with respect to some matrix norm) with $\ell$ invariant sets, then $T$ has $\ell$ eigenvalues close to~$1$, and the corresponding eigenvectors are close to those in~(\ref{eq:invariant_evec}). That means, in turn, if $T$ has eigenvalues close to one, we can identify regions (unions of boxes belonging to the same index set $\set{I}_i$) that the system is unlikely to leave within time~$\Delta$, i.e., they are \emph{almost invariant}~\cite{Dav82aa,GaSch98,DeJu99,DeWe04,GaSch06}.

\begin{figure}[ht]
\centering
\includegraphics[width=\textwidth]{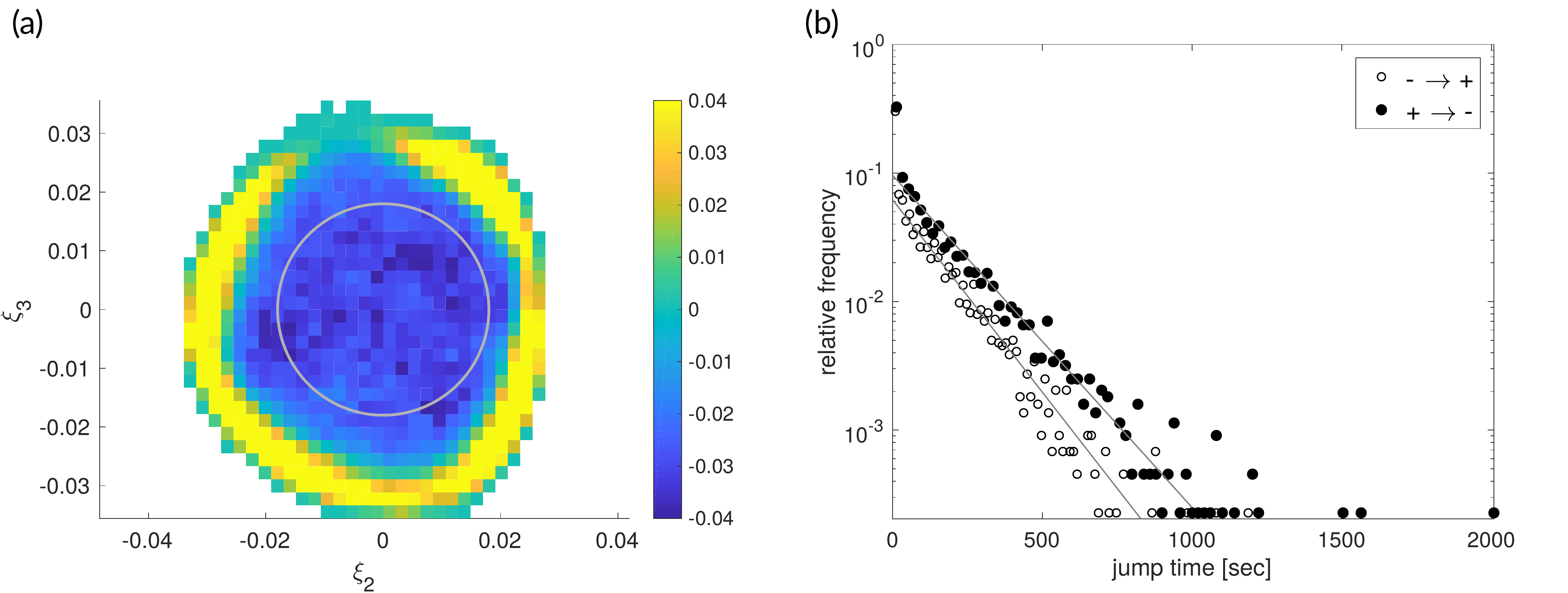}
\caption{Analysis of dataset D1. (a) Eigenvector $\rho_8$ at the largest subdominant purely real eigenvalue of the transition matrix constructed with $\ep=5\cdot 10^{-4}$, $\xi_{\bullet} = (\xi_2,\xi_3)$, $s=10$, and with an initial $32\times 32$ box covering of the bounding box of the data leading to~$p=754$. The color scale has been adjusted for better visual distinguishability of positive and negative regions. The circle represents the boundary in the embedding space between SRS and DRS as determined from fig.~\ref{fig:embedding_12}(d). (b) Switching time distribution between ``outer'' (yellow, positive) and ``inner'' (blue, negative) regions of the box partition. The gray lines are log-linear fits to the respective distribution up to jump times of~1000\,s. The switching time distributions seem to be exponential with mean switching times $t_{-\
\to\ +} = (145 \pm 5)$\,s and $t_{+\ \to\ -}=(168\pm 5)$\,s.
}

\label{fig:1003261_almostinv}
\end{figure}

%
%

In fig.~\ref{fig:1003261_almostinv}(a) we show the eigenvector of $T$ (obtained with $s=10$) corresponding to the largest 
real eigenvalue, $\lambda_8 = 0.496$. As $\tau \approx 3.45$\,s, the decay rate of this eigenvalue is
\[
\kappa = \frac{\log (\lambda)}{\Delta} = -0.020\,\mbox{s}^{-1},
\]
indicating transitions between the blue and yellow regions of the figure (positive and negative
parts of the eigenvector) happening on the order of timescale~$\kappa^{-1} \approx 50$\,s. 

As the sign structure of the eigenvector distinguishes regions well described by some level set of~$r_{\xi}$, this means that the largest almost invariance of the dynamics is in the radial direction. Comparison with fig.~\ref{fig:embedding_12} suggests that the dynamical feature found here approximately separates the SRS from non-SRS of the convection. Note that this is not the same as separating SRS from DRS, as this separation would be done by the gray circle in fig~\ref{fig:1003261_almostinv}, as obtained from fig.~\ref{fig:embedding_12}(d).

Computing expected lifetimes supports this observation. In \cite{WA11a} expected lifetimes of 53\,s
for the DRS and 270\,s for the SRS were measured, while both seemed to be exponentially distributed
random variables.  Fig.~\ref{fig:1003261_almostinv}(b) shows the empirical distribution of jump
times between the almost invariant sets. They seem exponentially distributed too, suggesting the
dynamics on this level of coarseness (i.e., only observing whether $r_{\xi}>c$ for some appropriate
constant~$c$) is Markovian. The expected transition time between the yellow and the blue areas are
($145\pm 5$)\,s (blue $\rightarrow$ yellow) and ($168\pm 5$)\,s (yellow $\rightarrow$ blue).
The numerical values in these two studies are of the same order of magnitude, however the discrepancy between them is not surprising. It stems from the fact that~\cite{WA11a} was measuring lifetimes of SRS and DRS, however there is a transition region between the two which does not belong to either. Our analysis assigns the transition region, the DRSs and some SRSs with smaller $r_{\xi}$ to one almost-invariant set (blue region), and the rest to the other.

\paragraph{Almost cyclic behavior.}

Let us consider an cyclic permutation $\sigma:\{1,\ldots,p\} \to \{1,\ldots,p\}$ with period~$p$ on the set of boxes. The associated transition matrix satisfies~$S_{ab} = 1$ if $b = \sigma(a)$, otherwise~$S_{ab}=0$. It is well known that the eigenvalues of $S$ are the $p$ roots of unity,~$\lambda_k = \omega^k$, where $\omega = e^{2\pi \imu/p}$ and~$\imu$ is the imaginary unit. The corresponding right eigenvectors~$\rho_k$ are 
\begin{equation} \label{eq:cyclic_evec}
\big(\rho_k\big)_{\sigma(a)} = \omega^k \big(\rho_k\big)_a.
\end{equation}
It follows that $S^p=\mathrm{Id}$, the identity, and thus a permutation induces a cyclic behavior. If all we know is $S$, we can deduce the dynamics from its eigenvectors by~(\ref{eq:cyclic_evec}). Note that every permutation matrix is also a stochastic matrix.

With a similar reasoning one can also relax the permutation requirement on $S$ in the following ways:
\begin{enumerate}[(a)]
\item
Let $S$ be a stochastic matrix with permuting blocks, i.e., there are $\set{I}_1, \ldots, \set{I}_\ell$ with~$\set{I}_1\cup\ldots\cup \set{I}_\ell = \{1,\ldots, p\}$ and a permutation~$\sigma:\{1,\ldots,\ell\} \to \{1,\ldots,\ell\}$ such that if $a \in \set{I}_i$ then $S_{ab} = 0$ for all $b\notin \set{I}_{\sigma(i)}$. This means that $S$ is a permutation viewed block-wise as given by the index sets~$\set{I}_i$. Then, $e^{2\pi\imu k/\ell}$ are among the eigenvalues of $S$, and the corresponding eigenvectors are constant on the $\set{I}_i$.
\item
If the matrix $S$ is stochastic, and close to a permutation matrix, then also its eigenvalues and eigenvectors are close to those of a permutation matrix. In this sense, one can speak of almost-cyclic behavior.
\end{enumerate}
To summarize, if the transition matrix $T$ shows eigenvalues close to those of a permutation matrix,
{\em i.e.,} eigenvalues close to the unit circle in the complex plane, then the time-series is expected to have an almost cyclic component in it~\cite{DeJu99}.

\begin{figure}[ht]
\centering

\centering
\includegraphics[width = \textwidth]{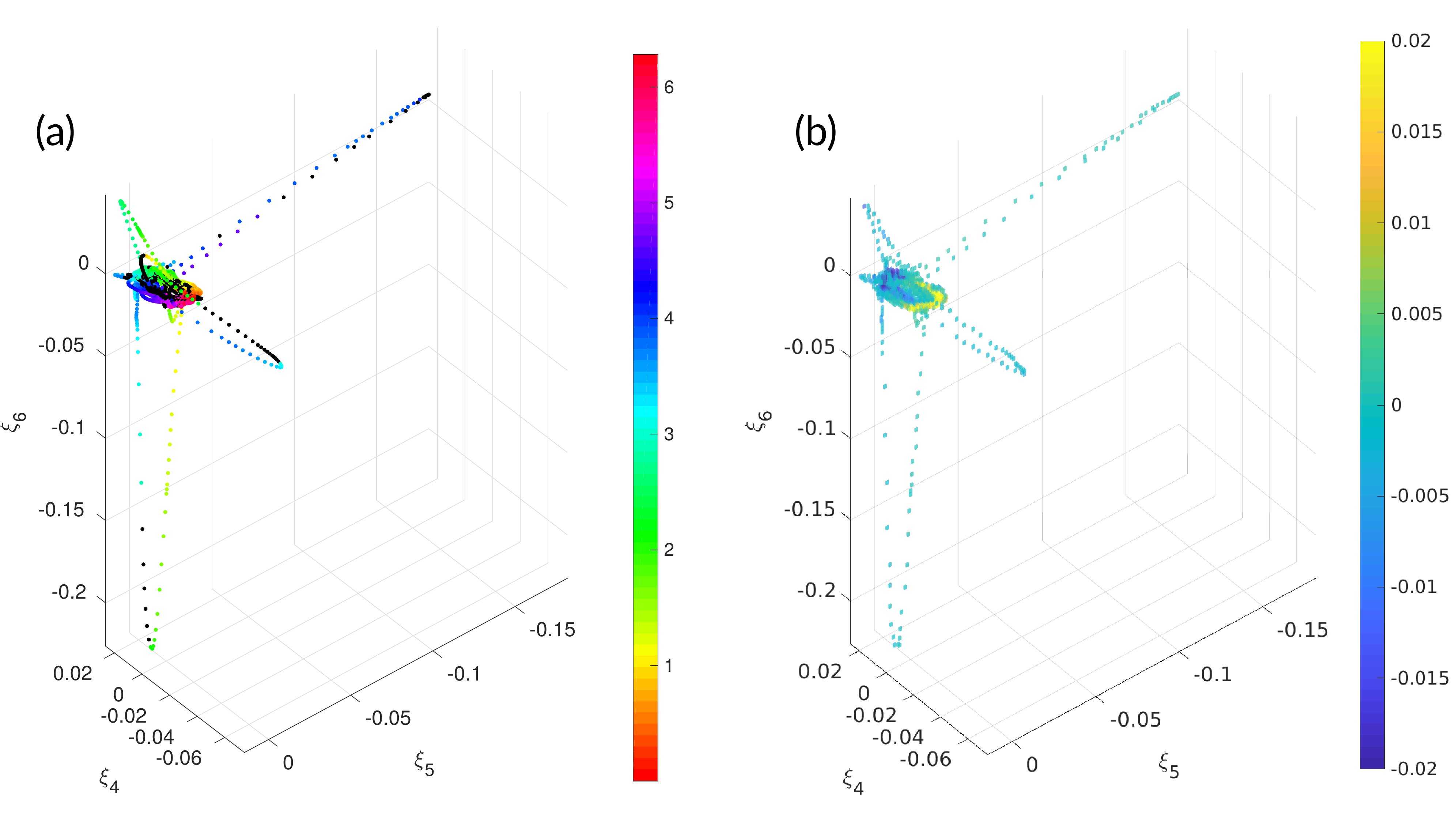}
\caption{(a) Embedding of D2 with $\ep = 7.5\cdot 10^{-5}$, coloring represents the orientation of the LSC
as measured in ~\cite{WA11d}. (b) Real part of the second eigenvector of the transition matrix on a $856$-box
covering with offset $s=5$, giving $\omega_0  = 1.17\cdot 10^{-2}$ and a period of $538$\,s.
}
\label{fig:1102271_3d_cyclic}
\end{figure}

Via transition matrix analysis one is able to find the hidden cyclic behavior even when considering
an embedding that does not map the data on a disc where an angular coordinate can readily be
identified. To show this, we use the automated $\ep$-selection procedure from
Appendix~\ref{app:opteps} for the data set D2, where the convection cylinder is rotating. This gives $\ep =
7.5\cdot 10^{-5}$. We subsample the data set, such that only every second measurement is used, and
``interpolate'' the remaining data points as described in Appendix~\ref{app:dmaps_interp}. For this
proximity parameter, the large but rare excursions dominate the geometry, and we choose
$\xi_{\bullet} = (\xi_4,\xi_5,\xi_6)$, where the disc-like geometry is not obvious. To compute the
transition matrix, we subdivide the bounding box of the embedded, three-dimensional data into
$128\times 128\times 128$ congruent boxes, by keeping only those that contain data points; leaving
us with~$p=856$ boxes. With the offset $s=5$ we compute the transition matrix $T$, and find the
second eigenvalue with decay rate
\[
\kappa = -0.0026 \pm 0.0116\imu.
\]
The real part of the corresponding eigenvector is shown in fig.~\ref{fig:1102271_3d_cyclic}(b). It
has negligibly small values on the excursing paths, and shows a sinusoidal pattern on the part of
the embedding space that can be attributed to the SRS; cf.\ fig.~\ref{fig:1102271_3d_cyclic}(a).
Application of $T$ turns the pattern of the eigenvector counterclockwise with period~$t_{\mbox{per}}
= \frac{2\pi}{\im{\kappa}} = 540$\,s. Note that in section~\ref{sec:orientdyn} we calculated for the
same dataset (D2) a rotation period of the structure of $2\pi/0.0122 $=515\,s.
The discrepancy is due to the embedding with different proximity parameters $\ep$ and due to the discretization error from the transition matrix computation on a finite partition. In summary, the transition matrix method reveals the cyclic behavior in the dynamics even for complicated embedding scenarios.

\paragraph{Transition matrix analysis summary.}

\begin{figure}[htp]
\centering
\includegraphics[angle=90,origin=c,height=0.7\textheight]{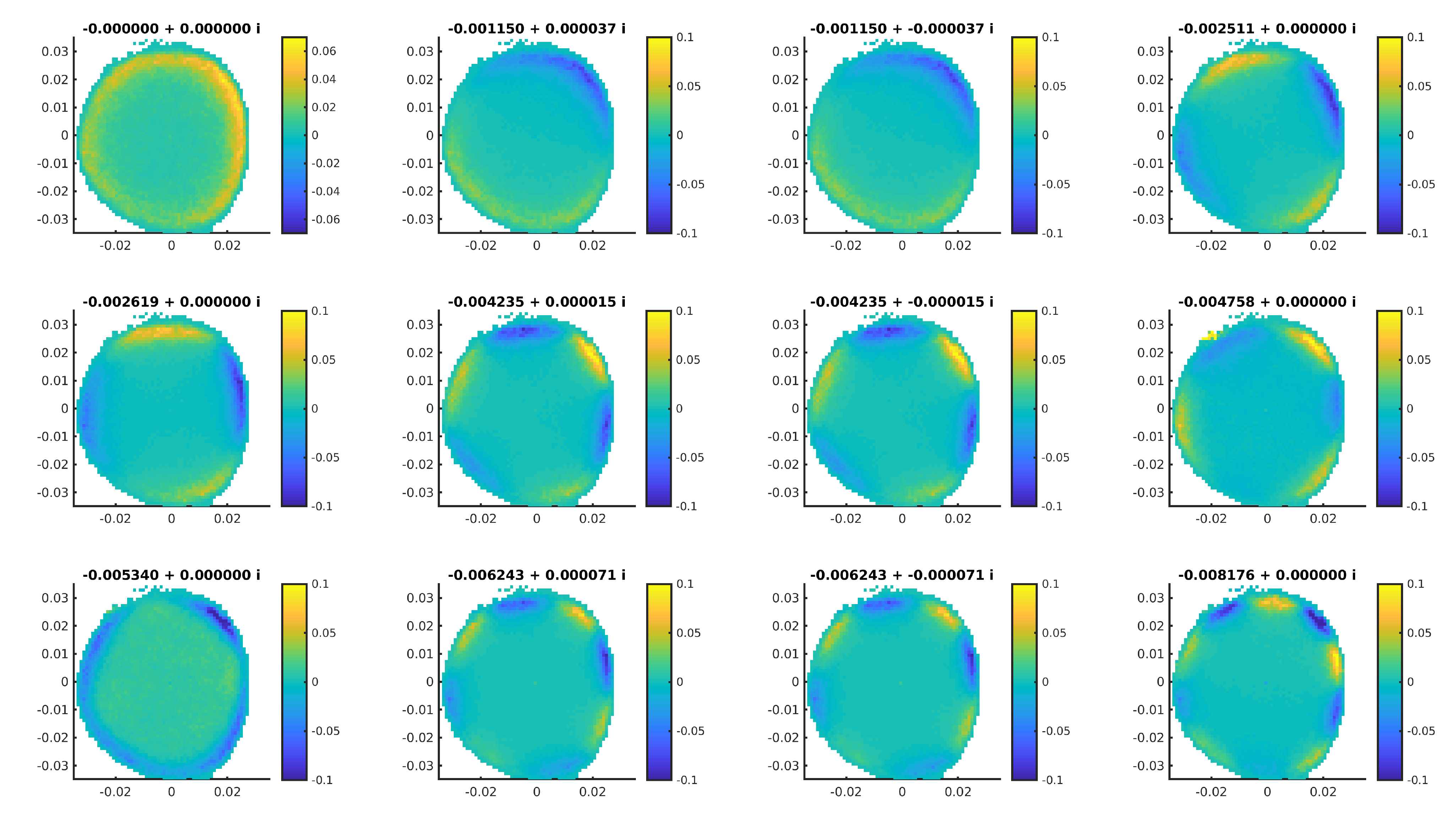}
\caption{Dataset D1. First 12 eigenvectors (left to right, row-wise from top to bottom) of the transition matrix constructed with $\ep=5\cdot 10^{-4}$, $\xi_{\bullet} = (\xi_2,\xi_3)$, $s=10$, and with an initial $64\times 64$ box covering of the bounding box leading to~$p= 2891$. The decay rates~$\frac{\log(\lambda_k)}{s\tau}$ are shown directly above the eigenvectors.}
\label{fig:1003261_allVs}
\end{figure}

\begin{figure}[htp]
\centering
\includegraphics[angle=90,origin=c,height=0.7\textheight]{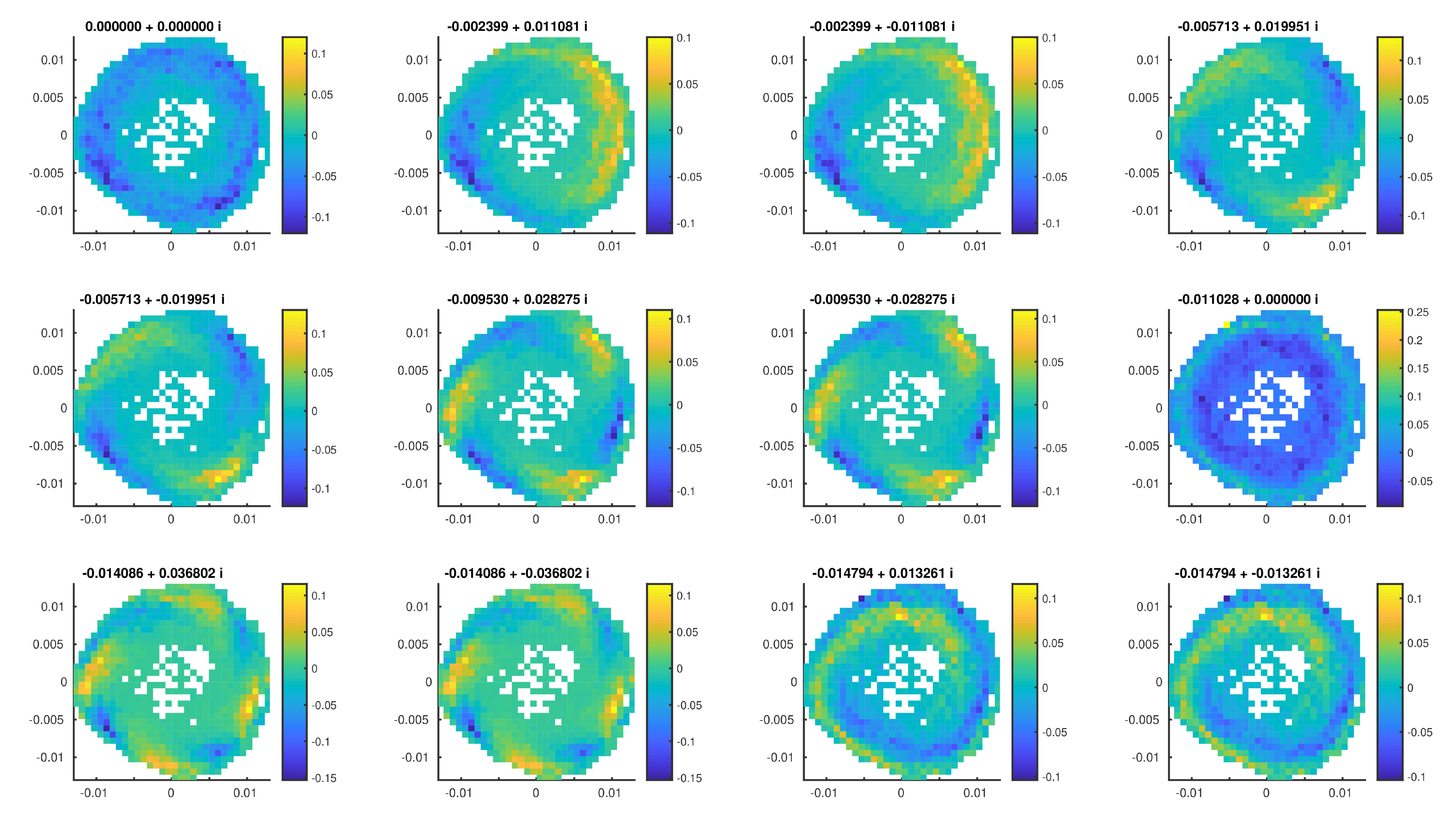}
\caption{Dataset D2. First 12 eigenvectors (left to right, row-wise from top to bottom) of the transition matrix constructed with $\ep=5\cdot 10^{-4}$, $\xi_{\bullet} = (\xi_2,\xi_3)$, $s=5$, and with an initial $32\times 32$ box covering of the bounding box leading to~$p= 714$. The decay rates~$\frac{\log(\lambda_k)}{s\tau}$ are shown directly above the eigenvectors.}
\label{fig:1102271_allVs}
\end{figure}

In figures~\ref{fig:1003261_allVs} and~\ref{fig:1102271_allVs} we depict the real parts of the first
12 left eigenvectors of the transition matrix for the data sets D1 and D2, respectively.
Their eigenvectors can be separated roughly in two classes: the first showing sinusoidal patterns in
the angular direction, and the second showing variation primarily only in the radial direction. This
suggests that---at least on long time scales---the dynamics in the orientation and the radial
direction are independent; otherwise we would expect a mixed-mode eigenvector, i.e., one that cannot
be written as a product~$\phi_{\mbox{ang}}(\theta)\phi_{\mbox{rad}}(r)$ of purely angular and radial modes.

If the dynamics were a pure noisy rotation of the orientation, the drift and diffusion coefficients $\omega_0$ and $D$ from section~\ref{sec:orientdyn} could be approximated from the eigenvalues of the transition matrix directly, as described in \autoref{app:noisyrot}. This works well for the angular frequency, as shown above for the data set D2, is however more defective for the diffusion coefficient.

\subsection{Comparison with other methods}

There is increasing activity on the dynamical analysis of flow fields; in particular the last decade
witnessed the development of different novel tools. We briefly discuss two such methods. The first one
seems to be the most widespread to date. The second is in spirit the closest to ours. Then
we draw a conceptual comparison with the theory of \emph{effective (or reduced) transfer operators}
and \emph{reaction coordinates}~\cite{BitEtAl18}, and discuss differences to Koopman Mode
Decomposition~\cite{RMBSH09}.

\paragraph{Dynamic mode decomposition.}

Dynamic mode decomposition (DMD), first introduced in~\cite{SS08}, was devised to reveal long-term
dynamical features of a system (eq.~(\ref{eq:orig_dyn})), when all that is available are observables~$z_1,\ldots,z_m$ sampled at a constant rate. With some offset $s$, building the data matrices
\[
X = \left( \begin{array}{ccc}
| & | & \\
z_1 & z_2 & \cdots \\
| & | & 
\end{array}
\right)
\quad\mbox{and}\quad
Y = \left(\begin{array}{ccc}
| & | & \\
z_{1+s} & z_{2+s} & \cdots \\
| & | & 
\end{array}\right),
\]
of the same size, one seeks a matrix (linear transformation) $A$ such that $AX \approx Y$, where this equality is solved in a least-squares fashion (by minimizing the Frobenius norm); i.e., $A = YX^+$ with $X^+$ being the pseudoinverse of~$X$.

In~\cite{williams2015data}, a connection of DMD and the so-called Koopman operator\footnote{For
functions $f:\set{X}\to\mathbb{C}$, the Koopman operator $\mathcal{K}^{\tau}$ associated with the
dynamical system $F^{\tau}$ is given by the linear mapping~$f\mapsto f\circ F^{\tau}$ in the
deterministic, and by~$f\mapsto \mathsf{E} [f\circ F^{\tau}]$ in the stochastic dynamical case.
Here, $\mathsf{E}[\cdot]$ denotes expectation. The Perron--Frobenius operator $\mathcal{P}^{\tau}$,
usually considered on the space $L^1$ of integrable functions, is its dual, and is uniquely defined
by $\int \mathcal{P}^{\tau} f\, g = \int f\,\mathcal{K}^{\tau}g$. It describes the propagation of
probability distributions under the (deterministic or stochastic) dynamics~$F^{\tau}$;
see~\cite{LaMa94}, for instance. For both of these operators we use the general name \emph{transfer
operators}.} has been revealed, which has been extended to the Perron--Frobenius operator
in~\cite{KlKoSch16}, which is dual to the Koopman operator. In particular, DMD converges to a
Galerkin projection of the Koopman (and, by a similarity transformation, to the Perron--Frobenius)
operator onto the space spanned by functions linear in the observable vector~$z$;
cf.~\cite{KlKoSch16}.

The main connection to our work here is that the transition matrix $T$ (eq.~(\ref{eq:transmat})) is a discretization of the Perron--Frobenius operator as well, as described in~\cite[Section 3.2]{KlKoSch16}. We expect this discretization to be superior to DMD, as it is a projection onto hundreds of boxes in our example cases, while DMD is a projection merely onto the 24-dimensional observable space. Nevertheless, an application of DMD with offset $s=5$ to the data set D2 gives a mode with decay rate $\kappa = -0.0021 \pm 0.0116\imu$, in good agreement with the above. The reason for this is that the complex cyclic mode can be well approximated by the linear functions of the observable; e.g., $\sum_{k=1}^8 e^{2\pi k\imu/8}((z)_k + (z)_{k+8} + (z)_{k+16})$, where $z = ((z)_1,\ldots,(z)_{24})^T\in \R^{24}$. However, applying DMD to the data set D1, we do not find any mode with decay rate close to $\kappa = -0.02$, which we attribute to the fact that the radial direction in our embeddings above---where the dynamics with this decay rate is happening---is a nonlinear function of~$z$.

Thus, DMD can capture dynamical features that can be characterized by linear functions of the observable, but is oblivious to other dynamical behavior.

\paragraph{Direct Koopman analysis by diffusion maps.}

Berry, Giannakis, and Harlim recently developed a method~\cite{BeGiHa15} with strong connections to the one we presented here. They approximate the so-called \emph{(stochastic) Koopman generator}~$\mathcal{L}$, i.e., the generator of the stochastic differential equations assumed to be underlying the dynamics that governs the data, directly on the diffusion maps eigenvectors~$\Xi_i$. To this end they use the approximation
\[
\mathcal{L}\Xi(z_j) \approx \frac{1}{\tau}\big(\Xi(z_{j+1}) - \Xi(z_j)\big)
\]
to obtain a converging (Galerkin) approximation as the number of dynamical samples grows. Our transition matrix $T$ can be seen as an approximation of~$e^{\tau\mathcal{L}}$ on a discretization of the domain of this operator; see~\cite{FrJuKo13} for connections between the transition matrix and discretization of the generator. The main difference lies in the discretization of the approximation space: while we are aggregating data points in a \emph{fixed} number of partition elements, the approximation~$\mathcal{L}\in\R^{m\times m}$ of~\cite{BeGiHa15} \emph{grows} in size with the size of the data set, and makes it thus numerically intractable at some point. It should be nevertheless remarked that by combining their method with the out-of-sample extension of diffusion maps, as used here, it is possible to compute more accurate (Galerkin) approximations of~$\mathcal{L}$ on a fixed tractable set of diffusion maps eigenfuctions~\cite{TGDW19}.

\paragraph{Effective transfer operators and Koopman Mode Decomposition.}

Effective (reduced) transfer operators arise in the context when the state of the system is only observed partially, through some non-linear observation function~$\xi: \set{X}\to\R^r$. They propagate distributions or observables $f_{\xi}$ that are functions of $\xi$, i.e., $f_{\xi} = \tilde{f}\circ \xi$ with $\tilde{f}:\R^r \to \mathbb{C}$, describing the effect of the dynamics as seen through the observation function~$\xi$, conditional to the system being in equilibrium. Thus, they are given by the conditional expectations
\begin{equation} \label{eq:effectiveTO}
\mathcal{P}^{\tau}_{\xi} f_{\xi}(x) = \mathsf{E}\left[ \mathcal{P}^{\tau}f_{\xi}(z) \,\big\vert\, \xi(z) = \xi(x) \right],
\qquad
\mathcal{K}^{\tau}_{\xi} f_{\xi}(x) = \mathsf{E}\left[ \mathcal{K}^{\tau}f_{\xi}(z) \,\big\vert\, \xi(z) = \xi(x) \right],
\end{equation}
where $\mathsf{E}[\,\cdot\, ]$ denotes expectation with respect to the invariant measure of the system. It is immediate that our transition matrix $T$, defined by~\eqref{eq:transmat} in section~\ref{sec:transmat}, is a discretization of $\mathcal{P}^{\tau}_{\xi}$ with observation function~$\xi = \xi_{\bullet}\circ h$. In fact, it can be seen as a Galerkin projection of this operator on the space spanned by piecewise constant functions over the boxes~$\set{B}_i$, cf.~\cite{KlKoSch16} for further details. Its transpose $T^T$ is an analogous approximation of the effective Koopman operator~$\mathcal{K}^{\tau}_{\xi}$, which is the adjoint of~$\mathcal{P}^{\tau}_{\xi}$. It is shown for reversible dynamics in~\cite{BitEtAl18} that the dominant timescales of the original system are still well retained in the system observed through~$\xi$, i.e., the dominant spectra of $\mathcal{P}^{\tau}$ and $\mathcal{P}^{\tau}_{\xi}$ are close, if the dominant eigenvectors $\phi_j$ of $\mathcal{P}^{\tau}$ are well parametrized by the observation function, i.e., there exist $\smash{\tilde{\phi}_j}:\R^r\to \mathbb{C}$, such that $\phi_j \approx \smash{\tilde{\phi}_j} \circ \xi$. This non-linear representation property is often present in molecular systems where the long time scales are connected to transitions between regions of state space (``reactions''), hence such a $\xi$ is called \emph{reaction coordinate}.

Koopman Mode Decomposition (KMD)~\cite{RMBSH09}---of which DMD is a particular data-based approxi\-mation---also considers the dynamics affecting a vector-valued observation function~$\xi$. It assumes that the observation function can be decomposed in the eigenfunctions of $\mathcal{K}^{\tau}$, i.e., $\xi(\cdot) = \sum_{k=0}^{\infty} v_j \phi_j(\cdot)$, where $v_j\in\mathbb{C}^r$ are the vector-valued coefficients of this decomposition (the Koompan modes). The evolution of the observation $\xi(x(t))$ of the true systems state $x(t)$ can then be described as
\[
\mathsf{E}\left[ \xi(x(t))\,\big\vert\, x(0)=x \right] = \mathcal{K}^t\xi(x) = \sum_{k=0}^{\infty} \lambda_j^{t/\tau} v_j\phi_j(x),
\]
or some suitable truncation of this series.

Given some observation function~$\xi$, both KMD and spectral analysis of the effective transfer operator are model reduction tools for the original, complex, possibly high-dimensional deterministic or random system.
While KMD aims at reconstructing the (expected) dynamics pointwise, spectral analysis of the effective transfer operator gives information about the dynamical processes with the slowest time scales.
Requirements for good performance of both methods are in some sense quite opposite: for KMD to perform efficiently, one requires $\xi$ to be  (componentwise) representable through linear combinations of a reasonable number of eigenfunctions of the Koopman operator, while the effective transfer operators require the dominant eigenfunctions of the full transfer operators $\mathcal{P}^{\tau}$ or $\mathcal{K}^{\tau}$ to be approximately non-linearly parametrizable by a low-dimensional observation function.
From the perspective of having a complex (chaotic) system at hand, it appears more suitable to ask for the long-term dominant statistical behavior of the system, as delivered by analyzing the effective transfer operators, than to consider single trajectories, as given by KMD. Also, once we have given a fair parametrization $\xi$ of the (approximate) attractor of a system, unless important dynamical processes happen in dimensions of the attractor not seen by this parametrization, the eigenfunctions of the full transfer operators are non-linear functions of~$\xi$.

\section{Summary and discussion}

In this paper, we have analysed geometric and dynamical features of turbulent Rayleigh--B\'enard
convection using the diffusion maps embedding approach. The state of the system was measured by the
temperature at 24 different locations in the sidewall of the convection cylinder, and the approach
embeds the unordered set of measurements into a space of chosen dimensionality.

We applied this embedding to the data of two different data sets, one where the convection cylinder
was at rest and another one where it rotates with a constant angular speed. We found that in both
cases the transformed data set forms a disc in the space spanned by the two dominant diffusion-map
coordinates $\xi_2$ and~$\xi_3$. We found that this disc represents the large-scale circulation
(LSC) in the convection container, with the orientation of the LSC being represented by the azimuthal
location on the disc, and the strength of the LSC being represented by the radial distance from the
disc center.   

We furthermore investigated the double and single roll states and their representation in the
$(\xi_2,\xi_3)$-embedding. We found that points corresponding to the DRS were located at the center
of the disc, while points belonging to the SRS were located at the perimeter of the disc. Analysing
the azimuthal dynamics of the embedded data shows a diffusive process on large time scales and a
ballistic process on short time scales. This behavior supports a model for the large-scale
circulation based on a Langevin equation suggested by Brown and Ahlers~\cite{BA08a}. 

While we clearly see the diffusive azimuthal motion of the LSC in our approach, we did not detect
any clear signatures of the torsional or the sloshing mode of the LSC (see
e.g.,~\cite{ZXZSX09,BA09}).  We believe that this is mainly due to the small aspect ratio of the cell
$\Gamma=0.5$, causing the SRS to be rather unstable with frequent switching between SRS and DRS. Due
to this erratic behaviour of the SRS, periodic torsional and sloshing oscillations are difficult to
detect. However, we do expect to observe the torsional mode for $\Gamma=1$
cells when the SRS is more stable and the torsional mode better detectable. 

If the embedding does not represent the bulk dynamical behaviour (because, e.g., large but rare
dynamical excursions dominate the geometry in state space), a transition matrix analysis in the
low-dimensional embedding space can reveal hidden large-scale dynamical features, e.g., cyclic
dynamics; shown in Figure~\ref{fig:1102271_3d_cyclic}.

The proposed approach belongs to the class of transfer (or Koopman) operator methods to approximate
properties of dynamical systems. While having direct connections to other methods of this
class~\cite{RMBSH09,williams2015data,BeGiHa15}, it is a novel composition of the so-called Ulam
discretization of transfer operators~\cite{Ulam60,Fr98,DeJu99,KlKoSch16} and the diffusion
maps~\cite{CoLa06,CoLa06geometric} manifold learning approaches to approximate a so-called effective
transfer operator~\cite{BitEtAl18}. We discussed accuracy and numerical cost in connection with
related methods. We expect the (spectral) convergence of our method towards the effective transfer
operator~$\mathcal{P}^{\tau}_{\xi}$ from~\eqref{eq:effectiveTO} in the correct limit of infinite
data points and vanishing-diameter box-covering.
This should follow by combining results from~\cite{vLBeBo08,
DeJu99} and~\cite[Appendix~C]{KlKoSch16}, under the assumption that the data lies exactly on a
finite-dimensional smooth manifold.

We note that many of the findings reported here, can also be calculated by using a classical
approach, where the amplitude and orientation of the LSC is determined from fitting a cosine
function to the thermistors at a given vertical position \cite{BA07a,FBA08,BA09,BA08a,WA11c}. While
this approach assumes already the existence of an LSC, the diffusion maps approach in contrast is suitable to
detect unknown structure in data in the first place. 

In summary, our approach was able to provide geometric intuition about the dynamical state space (in
a general sense, the stochastic ``attractor'') of the \emph{infinite-dimensional} Rayleigh--B\'enard
convection system in a cylinder. Furthermore, the long-term dynamical behavior on this space was
revealed, and found to be in good quantitative agreement with previous experimental results. While
these experimental results heavily relied on the knowledge of the physical space's geometry and
model assumptions, our analysis is not utilizing such knowledge at any point.  We believe that this
and similar analysis techniques can help to improve the understanding and reduced modeling of high-
or infinite-dimensional complex systems.

\section*{Acknowledgement} We acknowledge support by the Priority Program 1881 ``Turbulent
Superstructures'' of the German Research Foundation (DFG). We thank Michael Wilczek and J\"org Schumacher for valuable discussions.
\appendix

\section{More on diffusion maps}
\label{app:dmaps_appendix}

\subsection{Diffusion distance}
\label{app:diffusion_distance}

To understand in which sense does diffusion maps retain the geometric properties of the data manifold after embedding, let us recall that $P$ is a stochastic matrix and thus $P_{ij}$ can be viewed as the probability that a random walker jumps from the data point~$z_i$ to the data point~$z_j$.

Now define a \emph{diffusion distance} as follows:
\begin{equation}
D(z_i,z_j)^2 = \sum_k \frac{|P_{ik}-P_{jk}|^2}{\pi_k}\mbox{.}
\end{equation}
Here, $\pi = (\pi_1,\ldots,\pi_m)^T$ the stationary distribution of the Markov chain, and satisfies
\[
\pi_i = \frac{d(z_i)}{\sum_k d(z_k)}\mbox{, such that }\quad \pi_i P_{ij}=\pi_j P_{ji}.
\]
In this definition, the diffusion distance $D(z_i,z_j)$ is small, if the transition probabilities from $z_i$ are similar to the transition probabilities from $z_j$. This can be written as $P_{i-} \approx P_{j-}$, where $P_{i-}$ denotes the $i$-th row of~$P$.

Thus, we note that for
\[
\hat{\xi}_i:= P_{i-}
\]
we have
\[
\big\| \hat{\xi}(z_i)-\hat{\xi}(z_j) \big\|_{1/\pi}^2 = D(z_i,z_j)^2\mbox{,}
\]
{\em i.e.,} a weighted Euclidean distance in the space of distributions on the data set corresponds to the diffusion distance on the data manifold. The important observation in~\cite{CoLa06} is that this now can be reformulated using the eigenvalue decomposition of $P$ to yield
\begin{equation}
D(z_i,z_j)^2 = \sum_{\ell=1}^m \Lambda_{\ell}^2(\Xi_{\ell}(z_i)-\Xi_{\ell} (z_j) )^2 \mbox{.}
\end{equation}
This now means that with the embedding
\begin{equation}
\xi(z_i) :=
	\left( \begin{array}{c}
		\Lambda_1\Xi_1(z_i)\\	
		\Lambda_2\Xi_2(z_i)\\	
				\vdots			\\
		\Lambda_m\Xi_m(z_i)
	\end{array}
	\right)
\end{equation}
satisfies 
\begin{equation} \label{eq:diff_dist_equiv}
\big\|\xi(z_i)-\xi(z_j) \big\|^2 = D(z_i,z_j)^2,
\end{equation}
where $\|\cdot\|$ is the usual Euclidean norm.

Eq.~(\ref{eq:diff_dist_equiv}) holds only for the full $m$-dimensional embedding. Assuming that the eigenvalues $\Lambda_{\ell}$ decay sufficiently fast (after sorting them accordingly), we can represent the data points
sufficiently well by just a few eigenvectors, as in~(\ref{eq:diff_embedding}), {\em i.e.,} in a lower-dimensional space. Also if this is not the case, one still obtains a valuable parametrization of the data manifold by just a few (usually the number of eigenvectors to take is the dimension of the data manifold) selected eigenvectors, as the other eigenvector do not contain \emph{additional topological information} (these are so-called higher order harmonics, cf.\ Appendix~\ref{app:Bessel} for an example). Of course, the quantitative property~(\ref{eq:diff_dist_equiv}) is lost, but on a qualitative level one still obtains a low-dimensional one-to-one embedding of the data manifold.

Let us now consider the statement from section~\ref{sec:application_to_data}, claiming that a larger proximity parameter emphasizes the one-dimensional transient loops less in the embedding. Note that the data points on the transient arc are sparsely spaced along a one-dimensional line. This means, for a small $\ep$ there will be non-negligible transition probability $P_{ij}$ essentially only between neighboring points $z_i $ and~$z_j$. Thus, the random walker needs many steps to transition from one part of the arc to another. Meanwhile, for a larger $\ep$, transitions jumping over several neighbors are possible, speeding up transitions strongly. In the bulk, for both small and large values of~$\ep$ there are plenty neighbors present,  allowing for a fast diffusive spreading. Thus, the diffusion distance between points in the bulk stays low in $\ep$ is changed from large to small, while on the transient arc it becomes large. If the pairwise mutual distance between the arc points is large, they need to fill in more space in the embedding.

\subsection{Optimal choice of the proximity parameter $\ep$}
\label{app:opteps}

Here we will consider the task of automatically determining a ``good'' proximity parameter value~$\ep$ (sometimes also called \emph{bandwidth}) in the diffusion maps method. The ideas presented here stem from~\cite{CSSS08}, and have been later refined in~\cite{BeHa16}.

Let us consider the entries $K_{ij}(\ep) = \exp\left(-\ep^{-1}\|x_i-x_j\|^2\right)$ of the similarity matrix in the diffusion maps construction. If there are $m$ data points sampling the manifold~$\set{M}$, then, by interpreting the following sum as Monte Carlo approximation of an integral, we obtain
\begin{eqnarray*}
S(\ep) := \sum_{i,j} K_{ij}(\ep) &\stackrel{(\ast)}{\approx} \frac{m^2}{\mathrm{vol}(\set{M})^2} \int_{\set{M}}\int_{\set{M}} \exp\left( -\ep^{-1}\|x-y\|^2\right)\,dx\,dy \\
&\stackrel{(\ast\ast)}{\approx} \frac{m^2}{\mathrm{vol}(\set{M})^2} \int_{\set{M}}\int_{\R^d} \exp\left( -\ep^{-1}\|x-y\|^2\right)\,dx\,dy \\
&= \frac{m^2}{\mathrm{vol}(\set{M})} (2\pi\ep)^{d/2}.
\end{eqnarray*}

Here, on the one hand, $(\ast)$ works if $\ep$ is sufficiently large, such that the point cloud $\{x_i\}_{i=1}^m$ ``resolves'' the functions $\exp\left(-\ep^{-1}\|\cdot-x_j\|^2\right)$ properly, such that the Monte Carlo estimation is valid. On the other hand, $(\ast\ast)$ is a good approximation, if $\ep$ is sufficiently small, such that the integral $\int_{\set{M}}\exp\left(-\ep^{-1}\|\cdot-y\|^2\right)$ is well approximated by the same integral on the tangential space~$\R^d$ of $\set{M}$ at~$y$. For this, the ``Gaussian bell'' should be sufficiently localized, i.e., $\ep$ small. It is assumed, that in between, there is a sweet spot for the values of $\ep$ that the above approximations hold. It follows that
\begin{equation} \label{eq:epsloglog}
\log (S(\ep)) \approx \frac{d}{2}\log(\ep) + \log\left( \frac{(2\pi)^{d/2}m^2}{\mathrm{vol}(\set{M})}\right).
\end{equation}

We note the two limiting behaviors:
\begin{itemize}
\item
As $\ep\to 0$, we have $K_{ij}(\ep) \to \delta_{ij}$, the Kronecker delta. Thus, $S(\ep) \to m$.
\item
As $\ep\to \infty$, we have $K_{ij}(\ep) \to 1$, thus $S(\ep) \to m^2$.
\end{itemize}
In between these two extremes there should be a region of linear growth in the double-logarithmic plot $S(\ep)$ versus $\ep$, according to~(\ref{eq:epsloglog}). This is suggested to be determined by maximizing~$\frac{d \log(S(\ep))}{d\log(\ep)}$. The idea is that such an $\ep$ is neither too small (compared with the data point density), nor too large (compared with the diameter of the data point cloud). Note that the procedure also gives an estimate of the dimension of the manifold~$\set{M}$ through the slope computed in~(\ref{eq:epsloglog}).

\subsection{Out-of-sample extension of diffusion maps}
\label{app:dmaps_interp}

Once we have computed the diffusion map matrix $P$ and its eigenpairs $(\Lambda,\Xi)$ for a given fixed set of data points, $z_1,\ldots,z_m$, we would like to embed a new data point $z$ without repeating the whole diffusion map computation anew on the entire data set augmented by~$z$. Especially, we would like to avoid solving the numerically expensive eigenvalue problem.

To this end, let us write the eigenvalue equation $P\Xi = \Lambda\Xi$ of the diffusion map matrix in the following functional form,
\begin{equation} \label{eq:funct_EVP1}
\Xi(z_i) = \frac{1}{\Lambda} \sum_{j=1}^m p(z_i,z_j)\Xi(z_j),
\end{equation}
where $p(z_i,z_j) = P_{ij}$. The main idea of the ``interpolation'' is to use~(\ref{eq:funct_EVP1}) to evaluate $\Xi$ in an arbitrary point~$z$, cf.~\cite[Def.~2]{CoLa06geometric}. For this we need that the function $z \mapsto p(z,z_j)$---for fixed data points $z_1,\ldots, z_m$ and given $z_j$---can be evaluated at any point of the space, not just in data points~$z_i$. This is possible by simply repeating the construction in (\ref{eq:dmap1})--(\ref{eq:dmap3}) with replacing $z_i$ by~$z$. There, the summation in (\ref{eq:dmap1}) and (\ref{eq:dmap2}) is carried out over the original data points only. Thus, (\ref{eq:dmap3}) gives the values $p(z,z_j)$, and by (\ref{eq:funct_EVP1}) we set
\begin{equation} \label{eq:funct_EVP2}
\Xi(z) = \frac{1}{\Lambda} \sum_{j=1}^m p(z,z_j)\Xi(z_j),
\end{equation}
where the $\Xi(z_j)$ have already been computed in advance.

Naturally, one can vectorize this procedure for a whole set $\{\bar{z}_1,\ldots, \bar{z}_M\}$ of new data points. Then we need to compute the interpolating matrix~$\bar{P} \in \R^{M\times m}$ with $\bar{P}_{ij} = p(\bar{z}_i, z_j)$, and $\smash{\bar{\Xi} := ( \Xi(\bar{z}_1), \ldots, \Xi(\bar{z}_M) )^T }$ is obtained by $\smash{\bar{\Xi} = \frac{1}{\Lambda}\bar{P}\Xi}$.

\section{Eigenmodes of the Laplacian on a disc}
\label{app:Bessel}

To be able to better understand the diffusion-maps  embedding, we will briefly consider the eigenvalue problem of the Laplacian on the disc $\set{D} = \{(x,y)\in\R^2: x^2+y^2\le 1\}$ with homogeneous Neumann boundary conditions.

\begin{figure}[h]
\centering

\includegraphics[width = \textwidth]{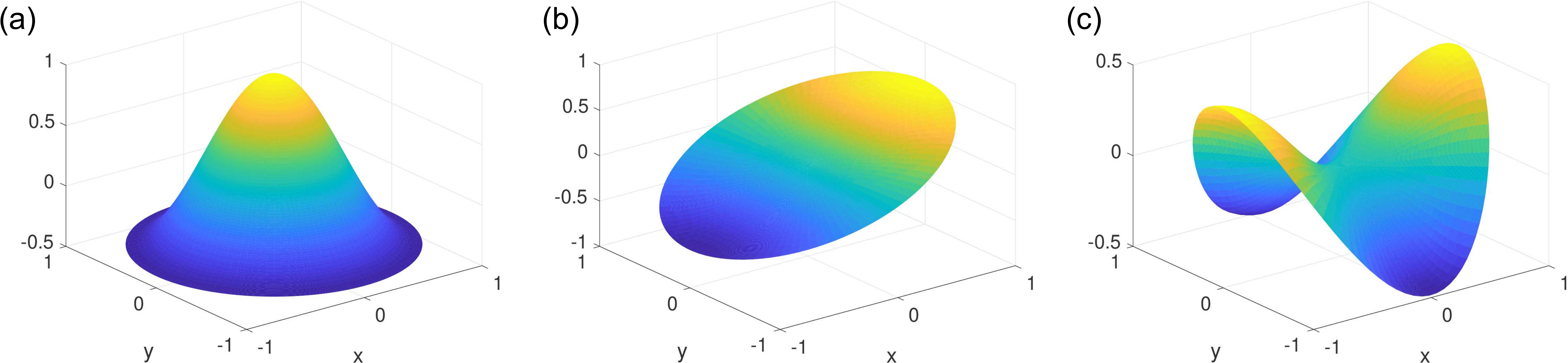}
\caption{Eigenfunctions of the Laplace operator. (a): $\psi_{0,1}$. (b): $\psi_{1,1}^{(1)}$. (c): $\psi_{2,1}^{(1)}$.}
\label{fig:LaplaceEfuns}
\end{figure}

It turns out~\cite[section~3.2]{GrNg13} that eigenfunctions are all separable in the sense that they can be written in polar coordinates~$(r,\phi)$ as
\begin{eqnarray*}
\psi_{n,m}^{(1)}(r,\phi) &= \cos(n\phi) J_n(\sqrt{-\lambda_{n,m}} r), \\
\psi_{n,m}^{(2)}(r,\phi) &= \sin(n\phi) J_n(\sqrt{-\lambda_{n,m}} r),
\end{eqnarray*}
where~$n \in \N_0$, $m\in\N$, $J_n$ is the $n$-th Bessel function of first kind, and~$\sqrt{-\lambda_{n,m}}$ is the $m$-th positive root of their derivatives~$J_n^\prime$. Due to the rotational symmetry in $\phi$, the eigenspaces for $n> 0$ are twice degenerate, and $\psi_{n,m}^{(1)}, \psi_{n,m}^{(2)}$ form a basis for them. If $n=0$, there is no $\psi_{0,m}^{(2)}$, because  the eigenspace is non-degenerate. The $\lambda_{n,m}$ are the associated eigenvalues. For instance:
\[
\lambda_{0,1} = -14.682,\quad
\lambda_{1,1} = -3.390, \quad
\lambda_{2,1} = -9.328.
\]
The corresponding eigenfunctions are shown in fig.~\ref{fig:LaplaceEfuns} as surface plots. If the disc has radius $R$ instead of one, the eigenvalues are scaled by~$\frac{1}{R^2}$.

\begin{figure}[h]
\centering
\includegraphics[width = \textwidth]{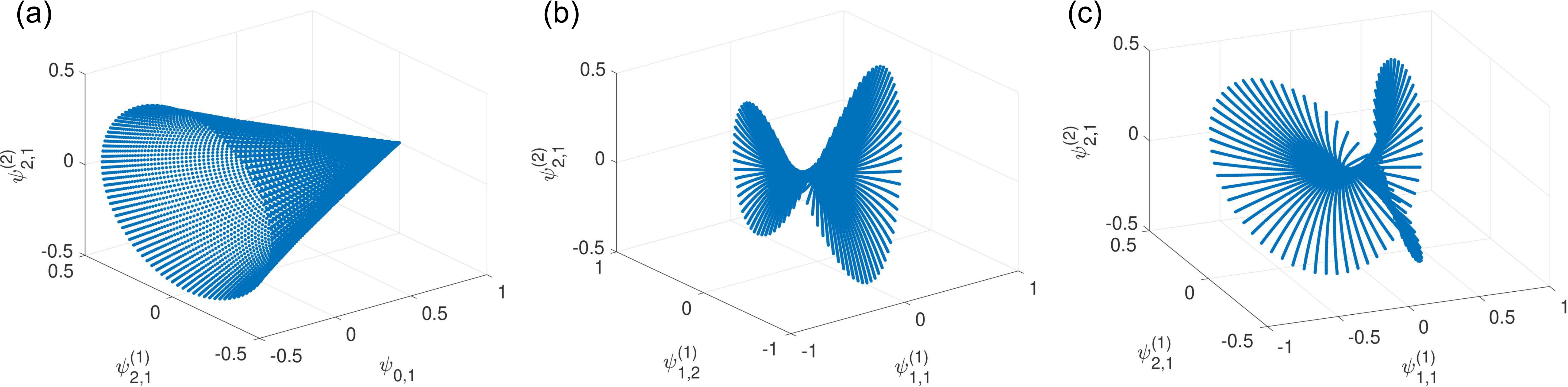}
\caption{Embeddings of the disc $\set{D}$ into $\R^3$ by different combinations of Laplacian
eigenfunctions. The cone-shaped (a), saddle-shaped (b), and figure-eight shaped (c) embeddings appear for the convection-roll data set as well, indicating a disc-like data manifold.}
\label{fig:LaplaceEmbed}
\end{figure}

Note that depending on the kernel function used in the diffusion maps method, one approximates different multiples of the Laplacian, i.e., in the appropriate sense
\[
\frac{P(\ep) - \mathrm{Id}}{\ep} \approx C\nabla^2,
\]
where for our kernel function $k(x,y) = \exp(-\ep^{-1}\|x-y\|^2)$ we have $C=\frac{1}{4}$, cf.~\cite[Theorem~25]{HAvL07}. Thus, the transformed eigenvalues $(\lambda-1)/\ep$ (or $\log(\lambda)/\ep$) of the diffusion map approximate $\frac{1}{4}\lambda_{n,m}$. As for high-dimensional data we cannot directly compare the diffusion map eigenfunctions with those of the Laplacian on a disc, an indication whether the data is an almost-isometric embedding of a disc in high dimensions is given by the following:

\begin{enumerate}[(i)]
\item The eigenvalues themselves depend on the radius of the circle, but their ratios should stay close to the ratios given by the $\lambda_{n,m}$ above.
\item Another indicator for a disc-like manifold is if the embedding by different diffusion-map eigenfunctions resembles an associated embedding by the eigenfunctions of the true Laplacian on a disc. Some of these are shown in fig.~\ref{fig:LaplaceEmbed}. Thy are to be compared with the embedding of the convection data in fig.~\ref{fig:1003261_pure}.  That these three-dimensional embeddings show two-dimensional surface, is an indication of degeneracy (non-linear interdependence) between the eigenfunctions. This is expected, as the manifold we consider is itself merely two-dimensional.
\end{enumerate}

\section{Noisy rotation on a circle}
\label{app:noisyrot}

Let us consider the SDE
\begin{equation}\label{eq:SDEnoisyrot}
d\theta(t) = \omega\,dt + \sqrt{D}\,dW(t),
\end{equation}
which describes a noisy uniform rotation on the circle with circumference $L$, that we identify with the (periodic) interval~$[0,L)$. The associated backward Kolmogorov equation reads as
\begin{equation}
\partial_t u = \frac{D}{2}\partial_{\theta\theta}u + \omega\, \partial_{\theta}u\quad\mbox{ for }u = u(t,\theta)\,.
\end{equation}
The associated eigenvalues (and eigenfunctions) are those of the right-hand side, which is a second order differential operator; called the \emph{generator}. Expressing the Kolmogorov equation for a Fourier basis with basis functions $\phi_k(\theta) = \exp\left(\imu \frac{2\pi k\theta}{L}\right)$, $k\in\mathbb{Z}$, we directly obtain that the basis functions are eigenfunctions at eigenvalues
\[
\lambda_k = -\frac{2\pi^2 k^2}{L^2}D + \imu\frac{2\pi k}{L}\omega.
\]
Thus, $\lambda_k, \lambda_{-k}$ are complex conjugate eigenpairs, and the corresponding eigenspace can be spanned by the real functions $\sin\left(\frac{2\pi k \theta}{L}\right)$ and~$\cos\left(\frac{2\pi k \theta}{L}\right)$.

The forward Kolmogorov (or Fokker--Planck) equation associated with~(\ref{eq:SDEnoisyrot}) reads as $\partial_t u = \frac{D}{2}\partial_{\theta\theta}u - \omega\, \partial_{\theta}u$, and has thus the very same eigenfunctions as the backward equation at complex conjugate eigenvalues, i.e., $\lambda_k^{fwd} = \bar{\lambda}_k^{bwd}$.

Thus, the noise and drift coefficients show up separately in the real and imaginary parts of the eigenvalues, and so we can estimate them from those.

\section{Azimuthal drift and diffusion as a function of the applied rotation rate}
We have explained in section~\ref{sec:orientdyn} how one can extract information of the azimuthal
orientation and their time dependence from the embedded data. We have shown that the azimuthal
orientation can be modeled with a Langevin equation that includes a constant azimuthal
drift~$\omega_0$. This drift is zero (or small for a single time trace) for a non-rotating system, but becomes important
under rotation.

\begin{figure}[htpb]
\centering
\includegraphics[width=0.8\textwidth]{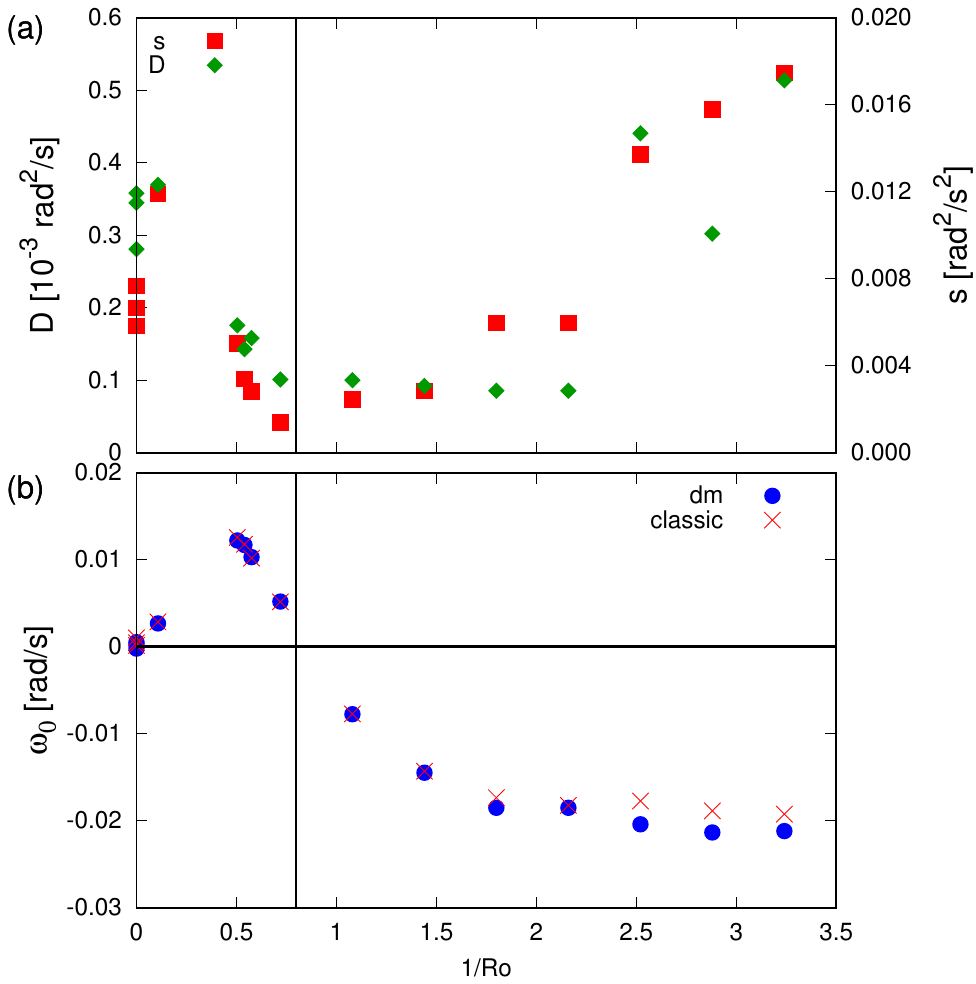}
\caption{Dynamics of the LSC as a function of the dimensionless rotation rate $1/\Ro$. (a) The
diffusion coefficient (green diamonds, left y-axis) and the ballistic coefficient (red squares,
right y-axis). (b) Drift $\omega_0$ as a function 1/\Ro\ calculated via diffusion maps as explained
above (blue bullets) and via the classical way (red x, see~\cite{WA11d}). One should note that since we cannot
determine the direction of the drift, i.e., the sign in front of $\omega_0$, we assigned the sign
from the classical analysis to it. Thus if the classical analysis shows cyclonic motion, we add
assumed that $\omega_0$ was positive, while we assumed negative $\omega_0$ when the classical
analysis shows anti-cyclonic motion. The black vertical line marks $1/\Ro=0.8$, where there onset of
heat transport enhancement due to rotation was observed (see \cite{WA11c}). 
}
\label{fig:diff_drift_rot}
\end{figure}

Fig.~\ref{fig:diff_drift_rot} shows the azimuthal diffusion $D$, the ballistic coefficient $s$ and
the average drift $\omega_0$ for different rotation rates (expressed dimensionless as the inverse
Rossby number 1/\Ro). Let us first have a look at the drift in fig.~\ref{fig:diff_drift_rot}b. Note that
using the diffusion maps embedding and calculating the azimuthal position from it, does not preserve
the sign of the azimuthal angle.
As a result we cannot
distinguish between cyclonic and anti-cyclonic motion of the flow structure. Thus, we assigned the
sign from the classical analysis to $\omega_0$. The result in fig.~\ref{fig:diff_drift_rot}b shows
that the in this way calculated azimuthal velocity (in fact these are not flow velocities but rather
velocities of the thermal structure) agree very well with the classically calculated values.
$|\omega_0|$ increase first with increasing 1/\Ro\, reach a maximum at around 1/\Ro$\approx0.5$ and
decrease after that. Another minimum is reached at around 1/\Ro$\approx 0.8$ and then $|\omega_0|$
increases again, with a negative sign. 

The diffusion rate $D$ and the ballistic coefficient $s$ also change as a function of 1/\Ro, as can
be seen in fig.~\ref{fig:diff_drift_rot}a. Interestingly, both values decrease initially with
increasing 1/\Ro, they reach a minimum at roughly 1/\Ro$\approx 0.8$ and increase for larger 1/\Ro.
Note that a change in the heat transport was observed at 1/\Ro=0.8. For faster rotation rates, the
vertical heat transport is enhanced compared to the non-rotating case.

We interpret the results found here, such that for sufficiently small rotation rates, the flow is
somehow stabilised, turbulent fluctuations are reduced or have less influence on the large-scale
circulation and thus the diffusion coefficient is reduced. The finding further support that at
1/\Ro=0.8 a transition occurs and that the fluid is in a different state for larger 1/\Ro. While it
is not clear what exactly happens at this point, there are evidences from previous studies that a
large convection roll is replaced by multiple vortices and columnar structures in which cold and
warm fluid is transported from the boundaries into the bulk (see e.g., \cite{WSZCLA10,SOLC11}). The diffusion of the structures is then no longer driven by the turbulent plume emission, but rather by the motion and interaction of the vortices.

\bibliography{bibliography}
\bibliographystyle{myalpha}

\end{document}